\documentclass[conference]{IEEEtran}

\usepackage{amsmath,amssymb,mathtools}
\usepackage{graphicx}
\usepackage{xcolor}
\usepackage{hyperref}
\usepackage{enumitem}
\usepackage{microtype}
\usepackage{dsfont}
\usepackage{bm}
\usepackage{makecell}
\usepackage{tabularx}

\usepackage{subfig}

\usepackage{dblfloatfix}

\usepackage[switch]{lineno}

\graphicspath{{figures/}}

\begin{document}
	
    \title{Optimality as a Generative Principle \\ for Network Structure}
	
	\author{
		\IEEEauthorblockN{Guram Mikaberidze}
		\IEEEauthorblockA{
			School of Computing\\
			University of Wyoming\\
			Email: gmikaber@uwyo.edu
		}
		\and
		\IEEEauthorblockN{Beso Mikaberidze}
		\IEEEauthorblockA{
			Muskhelishvili Institute of\\
			Computational Mathematics\\
			Georgian Technical University
		}
		\and
		\IEEEauthorblockN{Dane Taylor}
		\IEEEauthorblockA{
			School of Computing\\
			Department of Mathematics \& Statistics\\
			University of Wyoming
		}
	}
	
	\maketitle
	% \linenumbers
    
	\begin{abstract}
		Most real-world networks have evolved or been engineered to optimize some function, yet a unified framework for studying optimal networks across domains is lacking.
        We introduce \textsc{GradNet}, an AI-enabled framework that treats network topology as a continuously differentiable object, enabling the design and study of networks that optimize structural and dynamical objectives amenable to automatic differentiation under realistic constraints.
        We derive general optimality conditions, including an equimarginal principle for linear budgets, that make optimized networks analytically tractable.
        Canonical network features emerge spontaneously from constrained optimization: maximizing Kuramoto synchronization under coupling budgets yields sparse, bipartite, frequency-disassortative networks; minimizing social tension in opinion dynamics reproduces the factional split in Zachary's karate club; and maximizing communication capacity in spatial quantum networks under distance-dependent costs recovers minimum spanning trees.
        \textsc{GradNet} thus serves both as a network design tool scalable beyond $10^5$ nodes and as a scientific probe of structure-function relationships.
	\end{abstract}

	\section*{Introduction}
    
	Network science has largely focused on how connectivity structure governs network functionality \cite{albert2002statistical, newman2018networks}. In many systems, however, this relationship is inverted --- structure emerges from optimization toward some functional objective: engineered infrastructure is designed to minimize cost and maximize efficiency \cite{gastner2006shape}, river networks and other physical distribution systems are sculpted by energy dissipation minimization \cite{banavar1999size}, and biological networks, from circulatory system to gene regulatory circuits, are shaped by evolutionary and metabolic optimization \cite{west1997general, milo2002network}. A substantial body of work accordingly frames real networks as solutions to optimization problems, from highly optimized tolerance and efficient transportation \cite{banavar1999size, carlson2000highly, corson2010fluctuations} to neural wiring economy in brains \cite{kaiser2006nonoptimal, sporns2011non}. Yet, this literature remains fragmented across applications, objectives, and constraint types rather than constituting a unified field \cite{cancho2003optimization, seoane2015phase}.
    
    Developing only techniques to optimize networks, moreover, falls short of a deeper goal: understanding the properties of optimal networks themselves, including their emergent structural or dynamical mechanisms and the phase transitions that accompany them. Such questions remain largely unexplored relative to the mature study of dynamics on fixed networks and on random ensembles such as the Erd\H{o}s--R\'enyi, Watts--Strogatz, and Barab{\'a}si--Albert models \cite{erdos1959random, watts1998collective, barabasi1999emergence}. We argue that the study of optimal networks calls for a systematic formalization analogous to the variational and optimization formulations widely used across other fields (see Table~\ref{tab:variational_frameworks}) \cite{buchanan2007best}. Here we propose a differentiable optimization framework in which network topology and edge weights are treated as learnable variables subject to explicit structural constraints, enabling both the design of optimal networks and the systematic study of the structures that optimality produces.

    \begin{table}[!h]
		\centering
		\begin{tabular}{ll}
			\hline
			\textbf{Field} & \textbf{Variational / Optimization Principle} \\
			\hline
			Physics & Principle of least action \\
			Machine learning & Loss minimization \\
			Evolutionary biology & Fitness maximization \\
			Economics & Utility maximization \\
			Information theory & Rate--distortion optimization \\
			Statistics & Maximum likelihood estimation \\
			Control theory & Optimal control \\
			\textbf{Network science} & \textbf{Optimal network formulation, design \& study
            } \\
			\hline
		\end{tabular}		
		\caption{
            Across many scientific disciplines, variational and optimization principles serve as useful organizing frameworks alongside dynamical, empirical, statistical, and mechanistic approaches \cite{alexander1996optima,boyd2004convex,liberzon2011calculus}. In network science, despite a wealth of empirical and theoretical results, systematic optimization-based formulations remain comparatively underdeveloped and fragmented \cite{gastner2006shape, banavar1999size, west1997general, carlson2000highly, corson2010fluctuations, cancho2003optimization, seoane2015phase}.
        }
		\label{tab:variational_frameworks}
	\end{table}

	\begin{figure*}[!t]
		\centering
		\includegraphics[width=\textwidth]{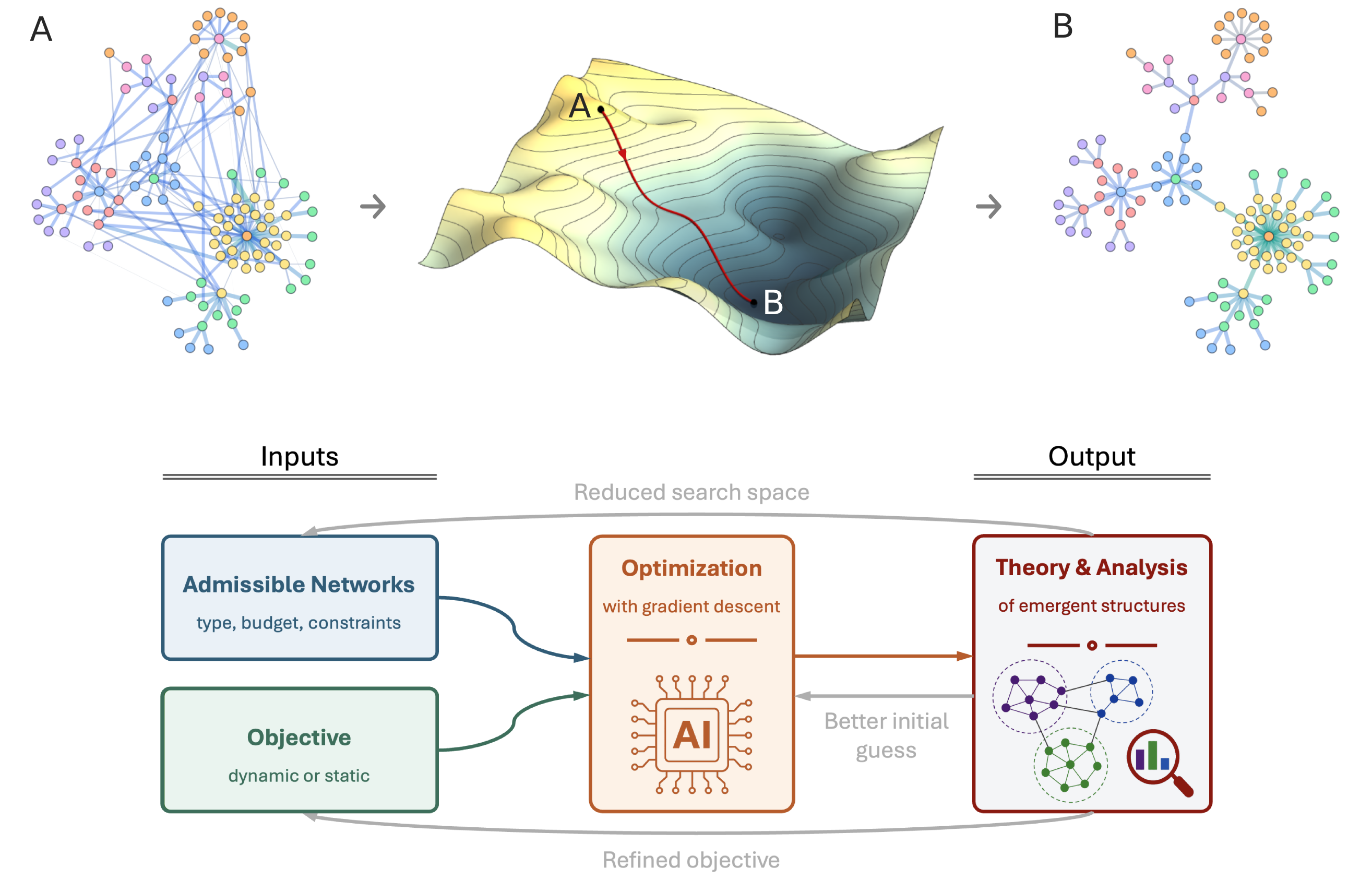}
        \caption{\textbf{Differentiable network optimization as a framework for design and discovery.}
        \textbf{Top:} Starting from an initial network (A), gradient-based updates move through a differentiable objective landscape toward an optimized network (B).
        \textbf{Bottom:} The problem is specified by a scalar objective and a set of admissible networks defined by properties and constraints, including directedness, construction budgets, and forbidden edges. The objective may be a direct function of network structure or may depend on a dynamical process unfolding on it. Gradient descent searches the admissible space using modern automatic-differentiation and machine-learning infrastructure. The optimized networks are subsequently analyzed to uncover emergent structural and dynamical principles, often in analytically tractable forms. The resulting theory can feed back into the optimization process by refining the objective, narrowing the search space, or providing improved initializations.}
        \label{fig:gradient_descent}
	\end{figure*}

    Computational efforts to optimize networks have proliferated, but remain tied to specific contexts: robustness \cite{schneider2011mitigation, zeng2012enhancing, chan2016optimizing, massei2024optimizing}, synchronizability \cite{motter2005network, donetti2005entangled, rad2008efficient, gorochowski2010evolving, taylor2016synchronization, mikaberidze2025emergent}, epidemic spreading \cite{boyd2004fastest, saha2015approximation, le2015met, xue2018distributed}, controllability \cite{becker2020network}, spectral quantities such as algebraic connectivity and effective resistance \cite{ghosh2006growing, ghosh2008minimizing}, and transportation \cite{gastner2006optimal, patwardhan2024symmetry}. Most target spectral surrogates, which allow each candidate mutation to be evaluated cheaply, or even perturbatively \cite{le2015met, xue2018distributed}, but which restrict the admissible objectives and decouple them from the dynamics they stand in for \cite{pecora1998master, wang2003epidemic, donetti2006optimal, skardal2014optimal}. The search itself is typically discrete and derivative-free: greedy rewiring \cite{schneider2011mitigation, zeng2012enhancing, chan2016optimizing, saha2015approximation, le2015met, xue2018distributed, patwardhan2024symmetry}, simulated annealing \cite{donetti2005entangled, rad2008efficient, gorochowski2010evolving}, or reinforcement learning over edge additions \cite{zhu2021network, li2023network}; convex formulations \cite{becker2020network, ghosh2006growing, ghosh2008minimizing} and hand-derived gradients \cite{taylor2016synchronization, boyd2004fastest} apply only when the objective permits. None of these differentiate through the objective, and the cost appears in scale: typically $10^3$--$10^4$ nodes, and tens to a few hundred for reinforcement learning. What is missing is a general differentiable parameterization of admissible networks, together with a characterization of the optima it produces.

    Realizing this vision demands a modeling framework general enough to span disparate objectives and constraints under a common formulation. Modern machine learning has produced a powerful optimization stack: automatic differentiation, GPU acceleration, and mature gradient-based optimizers \cite{baydin2018automatic}. This infrastructure, developed for training neural networks, can be repurposed to optimize network structure itself --- against static structural objectives or dynamical ones defined through processes unfolding on the network. This reframing positions network topology and edge weights as the continuously differentiable objects to be learned, thereby opening the door to variational theory to deepen our understanding of optimal networks. Our differentiable approach aligns both the design and theoretical study of optimal networks across applications.
    
    Thus motivated, we introduce \textsc{GradNet} (Fig.~\ref{fig:gradient_descent}) as a modeling framework for designing and studying functionally optimal weighted networks. One specifies the type and constraints of the network under question, such as its directedness and edge-construction budget, together with a scalar-valued objective function that evaluates the desired structural or dynamical property of the network. From these, \textsc{GradNet} builds a differentiable parameterization of the admissible networks, and then traverses it using gradient descent from arbitrary initial conditions toward minimizing the objective function. Built on the modern machine learning optimization stack, \textsc{GradNet} scales to networks of $10^5$ nodes, one to two orders of magnitude beyond the settings above. Beyond producing optimized networks, \textsc{GradNet} turns numerical design into a probe for discovery and theory: optimality is itself a strong mathematical condition, and the optimized structures \textsc{GradNet} reveals can be analytically tractable in both structure and dynamics.

    We demonstrate the breadth of this approach across diverse case studies spanning design, modification, and inference. A unifying theme runs through them: canonical structural features arise spontaneously from constrained optimization rather than being imposed as primitives. The most striking instance is sparsity, which materializes in several case studies even in the absence of any explicit sparsity penalty—a hallmark of real-world networks appearing here purely as a byproduct of optimization. Alongside this property, we find bipartition, minimum-spanning-tree–like backbones, and community partitioning whenever the objectives and constraints favor them. None of these features need be assumed in advance; each follows from constrained resource allocation under budgets, geometry, symmetry, and sign constraints. Across these cases, optimality serves as a generative principle for emergent structure, suggesting that many features long treated as descriptors may be better understood as consequences of constrained design.

    Numerical optimization alone does not explain why a structure is optimal, so we complement the framework with general optimality conditions. Under a linear budget with nonnegative weights, optimality reduces to an equimarginal principle: every realized edge returns the same marginal improvement per unit cost, and no absent edge could return more. More general and piecewise-differentiable objectives are handled by Lagrangian stationarity together with a candidate search over interior, boundary, and tie regions. These conditions make optimized networks analytically tractable, yielding closed-form optimal lattice weights (Eq.~\eqref{eq:l2_optimal_edgeweights_result}) and a proof that connected optima of the quantum-network objective are Euclidean minimum spanning trees (See \nameref{sec:methods}). 

    Classical generative models reproduce observed network features by building them into their rules. GRADNET instead asks which structures a given function demands under given constraints, so those features emerge as outcomes rather than assumptions. Optimality thus joins randomness and growth as a generative principle for networks, bringing to network science the variational framing long used in other fields (Table~\ref{tab:variational_frameworks}).

    \section*{Results}
    Below, we first present an overview of the \textsc{GradNet} framework, and derive analytical conditions for constrained network optimality. Next, we draw on these tools in six case studies (Table~\ref{tab:case_studies_summary}), spanning static and dynamic objectives; discrete- and continuous-time dynamics; and the construction, modification, and inference of networks.

    \begin{table*}[!t]
		\centering
		\small
		\setlength{\tabcolsep}{4pt}
		\renewcommand{\arraystretch}{1.8}
		\begin{tabularx}{\textwidth}{p{3.1cm} p{1.5cm} p{1.3cm} p{1.8cm} p{1.8cm} p{1.cm} p{2.4cm} p{2.8cm}}
			\hline
			\textbf{Case study} &
			\textbf{Directed or undirected} &
			\textbf{Modify or build} &
			\textbf{Modification signs} &
			\textbf{Edge-weight signs} &
			\textbf{Mask} &
			\textbf{Resource constraint form} &
			\textbf{Objective type} \\
			\hline
			
			Logical gates  &
			Directed &
			Build &
			Unconstrained &
			Unconstrained &
			N/A &
			N/A &
			Dynamic (discrete) \\
			
			Algebraic connectivity &
			Undirected &
			Build &
			Nonnegative &
			Nonnegative &
			2D grid &
			$B=\sum_{i j}A_{i j}$ &
			Static \phantom{(loss-func)} \\
			
			Zachary’s Karate Club &
			Undirected &
			Modify &
			Nonpositive &
			Nonnegative &
			N/A &
			N/A &
			Dynamic (continuous) \\
			
			Quantum internet &
			Undirected &
			Build &
			Nonnegative &
			Nonnegative &
			N/A &
			$B=\sum_{i j} C_{i j}A_{i j}$ &
			Static \phantom{(loss-func)}\\
            
			Kuramoto synchrony &
			Undirected &
			Build &
			Nonnegative &
			Nonnegative &
			N/A &
			$B=\sum_{i j}A_{i j}$ &
			Dynamic (continuous) \\
			
			Network reconstruction &
			Undirected &
			Build &
			Nonnegative &
			Nonnegative &
			N/A &
			$B=\sum_{i j}A_{i j}$ &
			Dynamic (continuous) \\
			
			\hline
		\end{tabularx}
    \caption{
        \textbf{Summary of the case studies and their optimization setups.}
        The columns correspond to the optimization space constraints and objective choices that define the problem. 
        The optimization space can consist of directed or undirected networks. 
        One can either seek optimal modifications of an existing network $A_{ij}=A^{(0)}_{ij}+\Delta_{ij}$, or build an optimal network from scratch (equivalent to modifying an empty network $A^{(0)}_{ij}=0$).
        One may impose restrictions on sign of the edge weight modifications $\Delta_{ij}$.
        Similar sign restrictions can be placed on the final edge weights $A_{ij}$.
        A mask can limit possible modifications to a subset of edges. 
        The resource constraint can be an arbitrary $\ell_p$ norm of cost-weighted edge weights $C_{ij}\Delta_{ij}$. In the spatial quantum-network case, $C_{ij}$ is taken proportional to the distance between nodes $i$ and $j$. Notice that when we are building a network from scratch $\bm A$ can be substituted for $\bm \Delta$. 
        Finally, the optimization objective may be a static function of the network structure, or it may depend on the discrete or continuous dynamical process.
    }
		\label{tab:case_studies_summary}
	\end{table*}

	\subsection*{\textsc{GradNet} Framework} \label{sec:framework_overview}
    
    The \textsc{GradNet} framework presented in Fig.~\ref{fig:gradient_descent} hinges on two essential specifications that together define the optimization problem: network constraints defining the optimization space and the optimization objective called the loss function. Here, we provide an overview of the framework and we defer the algorithmic details to \nameref{sec:methods}.

    The \textbf{optimization space} $\mathcal{A}$ is the set of admissible network adjacency matrices, represented as a constrained subset of $\mathbb{R}^{N\times N}$. Admissibility is determined by application-specific structural and resource constraints, including network directedness, allowable edge modifications, feasible edges, construction or modification budgets, variable costs, and restrictions on edge weights. \nameref{sec:methods} give the corresponding differentiable parameterization, exposed through \textsc{GradNet}'s simple interface, while further custom constraints can be baked into the definition of the loss function.

    The differentiable \textbf{loss function}, $\ell : \mathcal{A} \to \mathbb{R}$ quantifies the performance or fitness of the network for the desired objective. Importantly, the loss function may depend either directly on the network topology (e.g., graph diameter) or indirectly on the outcome of a dynamical process unfolding on the network. This dynamical process, in turn, can operate in either continuous or discrete time.

    Once the optimization space and the loss function are specified, \textbf{optimization} is performed via gradient-based methods to identify the network structure that minimizes the chosen loss ${\mathbf A}^* = \text{argmin}_{{\mathbf A} \in \mathcal{A}} \ell{({\mathbf A})}$. This is done by parameterizing the optimization space $\mathcal{A}$ using a differentiable encoding $g : \mathbb{R}^{N\times N} \to \mathcal{A}$ that maps freely tunable parameters $\mathbf P \in \mathbb{R}^{N\times N}$ onto all of $\mathcal{A}$  (the details and motivation of this encoding are given in \nameref{sec:methods}). Both $\mathbf P$ and $\mathbf A$ can be encoded densely or sparsely---the latter storing only the $E$ admissible edges at $O(E)$ rather than $O(N^2)$ cost, at the price of losing dense matrix algorithms such as diagonalization (see \nameref{sec:methods}). This yields a differentiable map from the tunable parameters $\mathbf P$ to the loss $\ell\big(g(\mathbf P)\big)$. Gradient descent can then traverse this loss landscape to uncover the optimal parameters $\mathbf P^*$. The optimal adjacency matrix follows as $\mathbf A^* = g(\mathbf P^*)$.

    The case studies below show how constrained optimization generates sparsity, symmetry, community separation, spanning-tree structure, bipartition, and monophily.

    \subsection*{Optimality Conditions} \label{sec:optimality}
    Beyond obtaining a numerically optimized network, a central scientific objective is to understand analytically why its structure is optimal (rightmost block in Fig.~\ref{fig:gradient_descent}). \textsc{GradNet} problems span diverse objectives and constraints, so no single analytical method covers them all; differentiable and piecewise-differentiable formulations, however, admit general stationary conditions and systematic procedures for identifying candidate optima. We develop these in three steps: a marginal-return condition for linear budgets, a candidate-enumeration procedure for smooth objectives under general constraints, and its extension to piecewise-differentiable objectives.

    \subsubsection*{Linear Budget and Nonnegative Edge Weights}\label{sec:opt_cond.A}
    Consider a nonnegative weighted network with an $\ell_1$ budget. Let $e$ index the admissible edges after symmetry and masking have been accounted for, let $a_e$ denote the weight of edge $e$, and let $c_e>0$ denote its cost per unit weight. The optimization problem is
    \begin{equation}
    \begin{aligned}
        \min_{\{a_e\}} \ell(\mathbf A), \quad
        \text{subject to} \quad
        a_e \geq 0, \quad
        m\sum_e c_e a_e = B.
    \end{aligned}
    \label{eq:linear_budget}
    \end{equation}
    Here, $B$ stands for the total budget. $m$ denotes the multiplicity of edges: $m=1$ if network is directed and $m=2$ if undirected. This convention counts each undirected edge twice, placing directed and undirected budgets on equal footing $B=\sum_{ij}A_{ij}$. Define the marginal improvement per unit cost as
    \begin{equation}
        \rho_e
        =
        -\frac{1}{m c_e}
        \frac{\partial \ell}{\partial a_e}.
    \label{eq:marginal_improvement}
    \end{equation}
    If two edges $e$ and $f$ both have positive weight, an infinitesimal transfer of budget from $f$ to $e$ changes the loss to first order by an amount proportional to $\rho_f-\rho_e$. At an optimum, no such transfer can decrease the loss. Therefore,
    \begin{equation}
        a_e^*>0
        \quad\Longrightarrow\quad
        \rho_e=\lambda,
    \label{eq:active_edge_condition}
    \end{equation}
    for some edge-independent constant $\lambda$. An admissible edge with zero weight must instead satisfy
    \begin{equation}
        a_e^*=0
        \quad\Longrightarrow\quad
        \rho_e\leq\lambda.
    \label{eq:inactive_edge_condition}
    \end{equation}
    This is an \textbf{equimarginal principle for edge weights}: all positive-weight edges provide the same marginal improvement per unit cost, whereas an absent edge cannot provide a larger improvement. The terminology is borrowed from environmental economics \cite{baumol1988theory}, and the principle arises in our context naturally whenever the budget is a linear combination of the edge weights. This principle is demonstrated most clearly in the ``\nameref{sec:algebraic}'' case study below.

    \subsubsection*{Smooth Objectives under General Constraints} \label{sec:opt_cond.B}
    More general constrained problems can be analyzed by considering the interior, relevant boundaries, and their intersections separately following the standard Lagrangian viewpoint of constrained nonlinear optimization~\cite{nocedal2006numerical}. Suppose the feasible set is described by equality constraints $h_r(\mathbf A)=0$ and inequality constraints $k_s(\mathbf A)\leq0$. We introduce the Lagrangian
    \begin{equation}
        \mathcal{L}
        \left(
            \mathbf A,\boldsymbol{\lambda},\boldsymbol{\mu}
        \right)
        =
        \ell(\mathbf A)
        +
        \sum_r \lambda_r h_r(\mathbf A)
        +
        \sum_s \mu_s k_s(\mathbf A),
    \label{eq:general_lagrangian}
    \end{equation}
    where the second sum includes only the inequalities saturated on the boundary under consideration. The corresponding stationary equations are
    \begin{equation}
        \frac{\partial \ell}{\partial a_e}
        +
        \sum_r
        \lambda_r
        \frac{\partial h_r}{\partial a_e}
        +
        \sum_s
        \mu_s
        \frac{\partial k_s}{\partial a_e}
        =
        0.
    \label{eq:general_stationarity}
    \end{equation}
    In the interior, no inequality is saturated and the second sum is omitted. On a selected boundary, its saturated constraints are treated as equalities, so their multipliers may have either sign. For each case, we solve the stationary equations, retain the candidates satisfying all original constraints, and compare their loss values. A simple illustration of this boundary perspective appears in the ``\nameref{sec:karate}'' case study below. Constraints not conveniently written in the form above can be handled through the same case-by-case search over feasible regions and boundaries.

    \subsubsection*{Piecewise-Differentiable Objectives}\label{sec:opt_cond.C}
    Objectives involving shortest paths, bottlenecks, and min/max operations are often only piecewise differentiable \cite{scholtes2012introduction}. Within a region where the active path, cut, bottleneck, or branch remains fixed, ordinary differentiation applies and the preceding procedure can be used. Tie surfaces (where two or more candidates attain the same optimal value, so the active one switches) separating such regions are analyzed directly as additional candidate sets, together with their relevant intersections and external boundaries. Feasible stationary candidates from all cases are then compared by their loss values. An example is given in the ``\nameref{sec:quantim_internet}'' case study.

	\subsection*{Recurrent Neural Networks Encoding Logical Gates}\label{sec:rnn}
	
	\begin{figure}[!t]
		\centering
		\includegraphics[width=0.99\columnwidth]{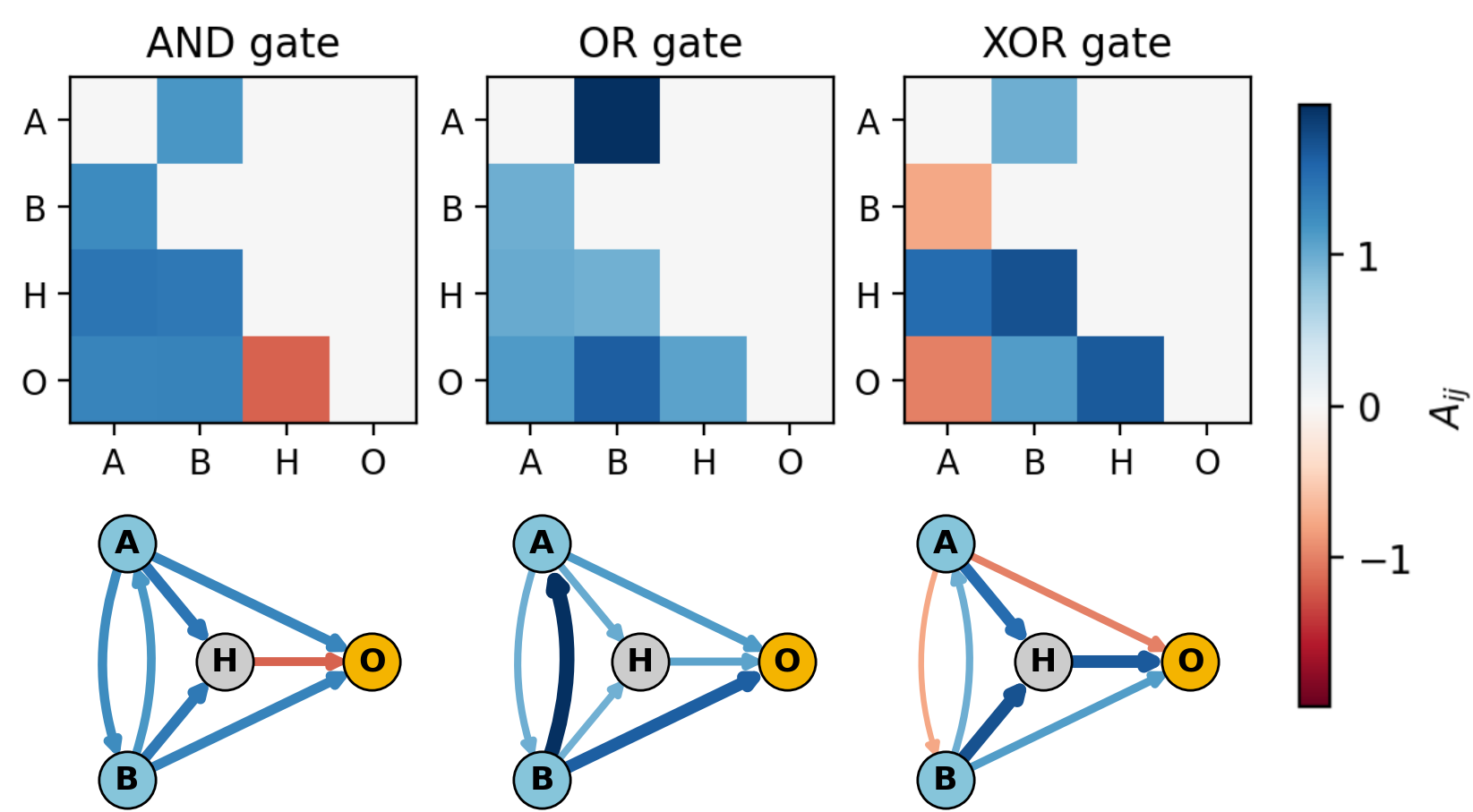}        
		\caption{\textbf{Networks that compute AND, OR, and XOR logical operations.} Nodes $A$ and $B$ serve as inputs, and after two iterations $\bm x(t+1) = \tanh\!\big(\mathbf A \bm x(t)\big)$ the result is written to node $O$. $H$ denotes a hidden node. The top row shows the optimized adjacency matrices, and the bottom row the corresponding network diagrams.}
		\label{fig:logical_gates}
	\end{figure}
	
    Our first case study trains toy recurrent neural networks to emulate logical gates, illustrating discrete-time dynamics on a network and invoking the historical role of XOR as a paradigmatic learning task~\cite{rumelhart1985learning}. We train small 4-node networks with adjacency matrix $A_{i j}$ to implement three operators estimating the logic gates AND, OR, and XOR. Nodes $\{A,B\}$ serve as inputs, the output is written to node $O$, and $H$ is a hidden node. The networks are directed with no self-loops, built from scratch with no budget and no sign constraints on edge weights. Hence, the parameterization of the optimization space simply maps arbitrary matrix $\mathbf P$ to one with a zero diagonal $A_{ij}=(1-\delta_{ij})P_{ij}$.
    
    % (ii) Loss
    The network evolves by two iterations of
    \[
      \bm{x}(t+1) = \tanh\!\bigl(\mathbf{A}\,\bm{x}(t)\bigr),
    \]
    which is message passing on a weighted directed graph, or equivalently two bias-free linear layers with $\tanh$ activation. The state $\bm{x}^{(i)}(t)=[x_A^{(i)},x_B^{(i)},x_H^{(i)},x_O^{(i)}]^\top$ is initialized from $\bm{x}^{(i)}(0)\in\{[0,0,0,0],[1,0,0,0],[0,1,0,0],[1,1,0,0]\}$, with nodes outside $\{A,B\}$ set to zero. The loss is the $\ell_2$ distance between the predicted output at $t=2$ and the appropriate target truth table $\{y^{(i)}\}$ for each logic gate:
    \[
      \ell(\mathbf{A}) = \sum_{i=1}^{4} \bigl|x_O^{(i)}(2) - y^{(i)}\bigr|^2.
    \]
    
    % (iii) Training
    A separate network is trained per gate from a zero initialization of $A_{i j}$. Starting from all zero edge weights though ensures that all the gradients of the loss function vanish. So to kick the training off the ground, we employ a small additive stochasticity $\epsilon=10^{-3}$ in our gradient descent.
    
    % (iv) Insight
    The optimized networks are shown in Fig.~\ref{fig:logical_gates}; in all cases, outputs round to the correct logical values. Notably, despite all edges being admissible, the learned networks are sparse. The mechanism is transparent: with only two message-passing steps, information travels along paths of length at most two, so the optimizer strengthens precisely those edges participating in length-two paths from $\{A,B\}$ to $O$, and ignores the rest. This illustrates how sparsity can arise from the dynamics and loss alone, without any explicit penalty. Allowing more hidden nodes or additional  message-passing rounds further reduces the loss, while the same path-length  argument predicts the resulting sparsity patterns.

	\subsection*{Structural Optimization: Algebraic Connectivity}\label{sec:algebraic}
	Our next example will optimize one of the most important global measures of a network, its algebraic connectivity \cite{fiedler1973algebraic, de2007old}, i.e., the second-smallest eigenvalue $\lambda_2$ of its associated Laplacian matrix with entries  $L_{i j} = \delta_{i j} \sum_k  A_{ik} -  A_{i j}$. 
    For an undirected network with $A_{ij}\ge 0$, $\lambda_1$ is always zero, while $\lambda_2$ is positive as long as the network is connected, its magnitude measuring how well connected it is.
    Improving  $\lambda_2$ for networks has been a target of widespread interest \cite{ghosh2006growing, mosk2008maximum, somisetty2024optimal}, since it can accelerate the convergence for dynamical processes over networks ranging from diffusion \cite{boyd2004fastest} and synchronization \cite{barahona2002synchronization} to consensus-based decision making in social networks \cite{bruggeman2018consensus}.
    Here, we employ \textsc{GradNet} to  computationally  reveal and analytically characterize optimally weighted lattices that maximize $\lambda_2$, highlighting how the framework can reveal structural patterns that allow seemingly difficult analytical challenges to become tractable.

    % \todo{(i) arch}
	We look for the best undirected network with the constrained coupling budget 
	\begin{equation} \label{laplacian_budget}
		B=\sum_{i j}A_{i j}.
	\end{equation}
	It turns out that in general, optimal resource allocation with this objective produces a completely and uniformly connected graph. Hence, as a more interesting demonstration, we restrict the network to an $n\times n$ 2D square lattice.
	
	\begin{figure}[!t]
		\centering
		\includegraphics[width=0.99\columnwidth]{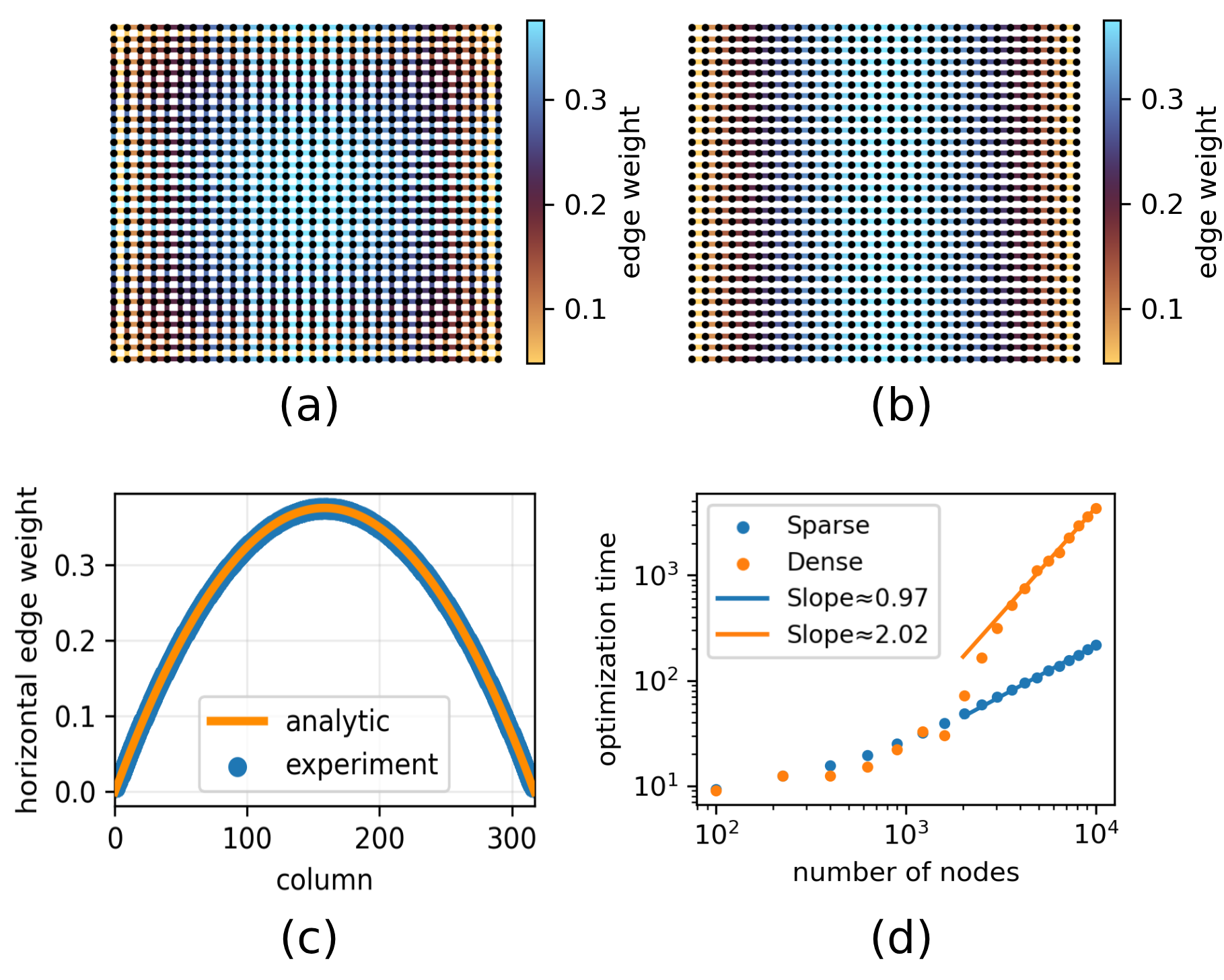}		
		\caption{\textbf{Optimal square lattice maximizing the algebraic connectivity}, i.e., the second Laplacian eigenvalue $\lambda_2$, under the constrained budget in Eq.~\eqref{laplacian_budget}. 
			(a) Heatmap of the optimal edge weights for a weighted $30\times 30$ square lattice. 
			(b) Horizontal edge weights plotted separately, revealing that they depend only on horizontal position (an analogous statement holds for vertical edges). 
			(c) Comparison between the analytical optimum (Eq.~\eqref{eq:l2_optimal_edgeweights_result}) in orange and the numerical optimum for $317\times317$ network, i.e., $N\approx10^5$ nodes, optimized on an NVIDIA L40S~GPU for 2.7 hours. 
			(d) Optimization time (seconds) versus number of nodes $N$ for a fixed number of optimization steps ($50$) of the power method, comparing sparse and dense encodings on an M2 chip MacBook. Linear fits to the final eight data points indicate $O(N)$ scaling for the sparse encoding and $O(N^2)$ scaling for the dense encoding.
		}
		\label{fig:l2_variants}
	\end{figure}

    % \todo{(ii) loss, (iii) train}
    Since we seek to maximize $\lambda_2$, we minimize the loss $\ell(\mathbf{A}) = -\lambda_2$. 
    This task serves as an excellent scenario for contrasting dense and sparse encodings of the adjacency matrix, revealing the advantages and disadvantages of each.
    
    For the densely encoded network, we compute eigenvalues using the standard dense eigensolver \texttt{torch.linalg.eigvalsh} native to \textsc{PyTorch}, whose LAPACK backend relies on QR/QL methods and has time complexity $O(N^3)$ \cite{anderson1999lapack}. The result of this optimization is shown in Fig.~\ref{fig:l2_variants}(a). For a sparsely encoded network, we compute $\lambda_2$ using a modified power method detailed in \nameref{sec:methods}.

    % \todo{(iv) discovery}
    Examination of the optimal edge weights in Fig.~\ref{fig:l2_variants} reveal that edges near the center play a critical role and should be prioritized with more resources. Moreover, one can observe that the solution has a particular symmetry: 
    the weights of vertical edges are independent of their horizontal positions, and similarly, the weights of horizontal edges are independent of their vertical positions (see Fig.~\ref{fig:l2_variants}(b)). 
    Closed-form expression of graph Laplacian matrix eigenvalues are generally unavailable, and this remains true for square lattices with arbitrary edge weights.
    But, as discussed above, the optimality is a significant simplifying condition allowing much deeper analysis.

    We label the nodes of an $n \times n$ square lattice by $(i,j)$. Then, let $a_{i j}$ denote the weight of the horizontal edge between $(i,j)$ and $(i,j+1)$, and $b_{i j}$ the weight of the vertical edge between $(i,j)$ and $(i+1,j)$. At the optimum, the budget constraint requires any marginal increase in one edge weight to be offset by a compensating decrease elsewhere, so no such reallocation can improve $\ell$. So, the optimal $\{a_{i j}\}$ and $\{b_{i j}\}$ are obtained by the equimarginal principle Eq.~\eqref{eq:active_edge_condition}:
    \begin{equation}
        \frac{\partial \ell}{\partial a_{i j}}
        = \frac{\partial \ell}{\partial b_{i j}}
        = c \quad \text{for all } i, j,
    \end{equation}
    for some constant $c$: the marginal return on $\ell$ is equal across every edge. Combining this observation with the derivation in \nameref{sec:methods}, we obtain the closed-form solution for
	\begin{equation} \label{eq:l2_optimal_edgeweights_result}
		a_{ik}=b_{kj} = \frac{3}{2}\frac{B}{(n^2-1)n^2} k(n-k).
	\end{equation}
	
	This expression exhibits close agreement with numerical optimization results, as demonstrated in Fig.~\ref{fig:l2_variants}(c).

    Figure~\ref{fig:l2_variants}(d) shows the comparison of sparse and dense encodings. The optimization time scales as $O(N)$ for sparse encoding and $O(N^2)$ for dense encoding. In general, dense encoding stores and moves $O(N^2)$ elements, but in exchange grants access to a much larger toolbox of matrix algorithms, including matrix inversion, diagonalization, and standard spectrum computation. Sparse encoding inverts this trade-off: it stores and moves only $O(E)$ non-zero matrix elements, where $E$ is the number of edges. It offers a substantial boost in speed and memory efficiency, but forfeits these dense-dependent operations, forcing us to resort to approximate methods instead.

	\subsection*{Empirical Network Enhancement: Zachary’s Karate Club}\label{sec:karate}
	\begin{figure}[!b]
		\centering
		\includegraphics[width=0.99\linewidth]{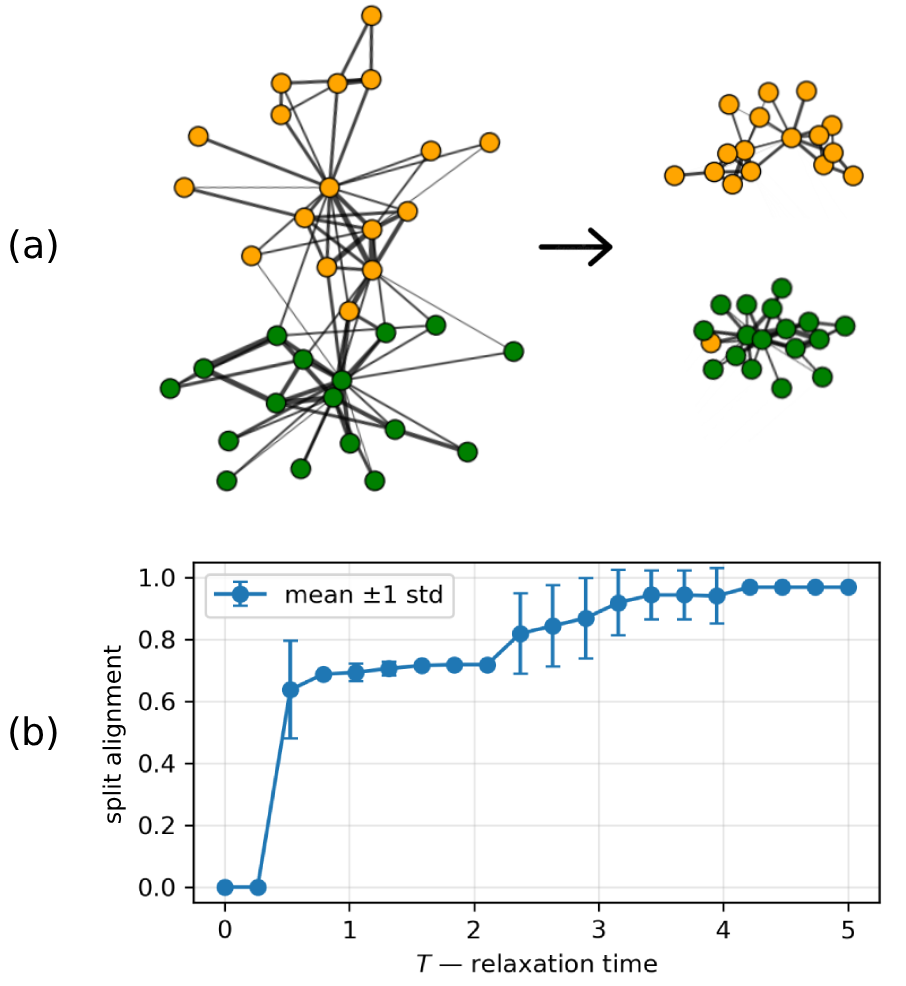}
        \caption{
            \textbf{Social tension drives Zachary's Karate Club to fragment.}
            \textbf{(a)} The original 34-member friendship network (left) and the network obtained by minimizing the social tension $\mathcal{D}(T)$ after opinion relaxation for time $T=10$ (right).
            Optimization separates the network into two connected components.
            Node colors indicate the clubs that members joined after the split; the inferred partition agrees with the empirical division for all but one member.
            \textbf{(b)} Split alignment---the fraction of members correctly assigned to their post-split club---as a function of the opinion-relaxation time $T$.
            For short relaxation times, pruning is driven primarily by local disagreement near the two opinion leaders and fails to recover the observed post-split division; as $T$ increases, opinions propagate through the network and the optimized partition approaches the empirical split.
            Error bars indicate the mean $\pm$ one standard deviation among $10$ independent optimizations.
		}
		\label{fig:zachary}
	\end{figure}

    This example illustrates that the framework can be used to study naturally occurring optimization processes. Here, we examine a well-studied scenario in which a social network adapted over time to resolve a conflict: Zachary's Karate Club \cite{zachary1977information}, a canonical benchmark in network science. It is a classic social network of 34 members whose friendships were recorded by Wayne Zachary in the 1970s, and a conflict between the instructor and the club president eventually caused the club to split into two separate clubs.
    
    This problem has been studied through the lens of relationship evolution, whereby the reduction of social tension or the restoration of structural balance leads to a two-faction organization centered around the instructor and the president \cite{jia2016coevolution, gorski2023general}. We model the evolution of social ties by minimizing social tension in the network: the two leaders hold fixed, opposing opinions that diffuse through the network. Friendships then rewire to reduce social tension, favoring ties between members with similar opinions, which subsequently splits the graph into two components.
    
	We model opinion exchange by Laplacian diffusion, a standard linear model of continuous opinion dynamics \cite{baumann2020laplacian}
	\begin{equation}\label{diffusion_dynamics}
		\bm{\dot o}(t) = -\,\mathbf{L}\bm{o}(t).
	\end{equation}
	Here $L$ is the Laplacian matrix. The opinions of the president and the instructor are fixed $o_1(t)=1,\;o_N(t)=-1$. 
    Everyone else is initially indifferent $o_i(0)=0$ for $1<i<N$.
    Following related disagreement and social-tension measures in opinion dynamics \cite{galbrun2017finding, musco2018minimizing}, we measure the instantaneous social tension by the normalized edge-disagreements
	\begin{equation}\label{social_tension}
		\mathcal{D}(t) = \frac{\sum_{i,j} A_{i j}\,\big(o_i(t) - o_j(t)\big)^2}{\sum_{i,j} A_{i j}}.
	\end{equation}

    % \todo{(i) structure}
	We initialize the \textsc{GradNet} optimization at the initial network Zachary measured, and optimize it from there. The final edges are restricted to be non-negative, and the allowed modifications are non-positive, to emulate breaking friendships. 
    
    % \todo{(ii) loss}
    The loss function $\ell=\mathcal{D}(T)$ is defined as the social tension after some time $T$ of the opinion diffusion. We assume that opinions diffuse faster than the friendships break. 
    This is a natural assumption given the causal relationship between the two processes, yet we explore the exact effects of $T$ below. 

    % \todo{(iii) optimization}
    We set $T=10$ and gradually prune the weighted network in order to minimize the loss function. For each iteration of gradient descent, we simulate the opinion diffusion, measure the instantaneous social tension at time $T$, and prune the network in an optimal manner to reduce $\ell(\mathbf A)$. Optimization naturally leads to a disconnected network, where the final stationary opinions within each connected component are constant, resulting in zero social tension. After this, the loss cannot be further improved and we terminate the optimization. This is a boundary optimum: the constraints \(A_{ij}\geq 0\) are active on the pruned edges, and the attained value \(D(T)=0\) is the global lower bound of the loss.

    % \todo{(iv) Analysis}
    The result of this pruning is shown in Fig.~\ref{fig:zachary}(a). The edges represent the friendships after the social tension has been minimized. You can see that the network has split into two connected components. The node colors, on the other hand, indicate the observed post-split club membership reported by Zachary. The network split produced by the opinion-diffusion model agrees with the observed post-split club membership for all but one member, node 9. Notably, Zachary reported that node 9 was a weak supporter of the president but nevertheless joined the instructor’s club because doing otherwise would have required abandoning his existing rank and beginning again in a different style of karate \cite{zachary1977information}. This example shows how clustering can emerge as a result of social pressure.

    Finally, it is interesting to discuss the effects of the time variable $T$. Let us note that if we give opinions very little time to relax $T\ll 1$, the opinions will have not spread through the network significantly, and the social tension $\mathcal D(T)$ will be mostly the result of the disagreements between the opinion leaders with $o(T)\approx \pm 1$ and their neighbors $o(T)\approx0$. Hence pruning will target these edges, isolating the opinion leaders, and classifying all club members incorrectly (into neither of the two post-split clubs). On the other hand, we have seen above, that if we wait for diffusion to relax $T\gg 1$, the procedure produces a split with high agreement with the observed post-split club membership. Figure~\ref{fig:zachary}(b) shows how this transition occurs as a function of $T$, suggesting that opinion dynamics in real social networks happens on a sufficiently faster time-scale than the evolution of social ties.

	\subsection*{Spatially Embedded Networks: Quantum Internet}\label{sec:quantim_internet}

	\begin{figure}[!b]
		\centering
		\includegraphics[width=0.99\linewidth]{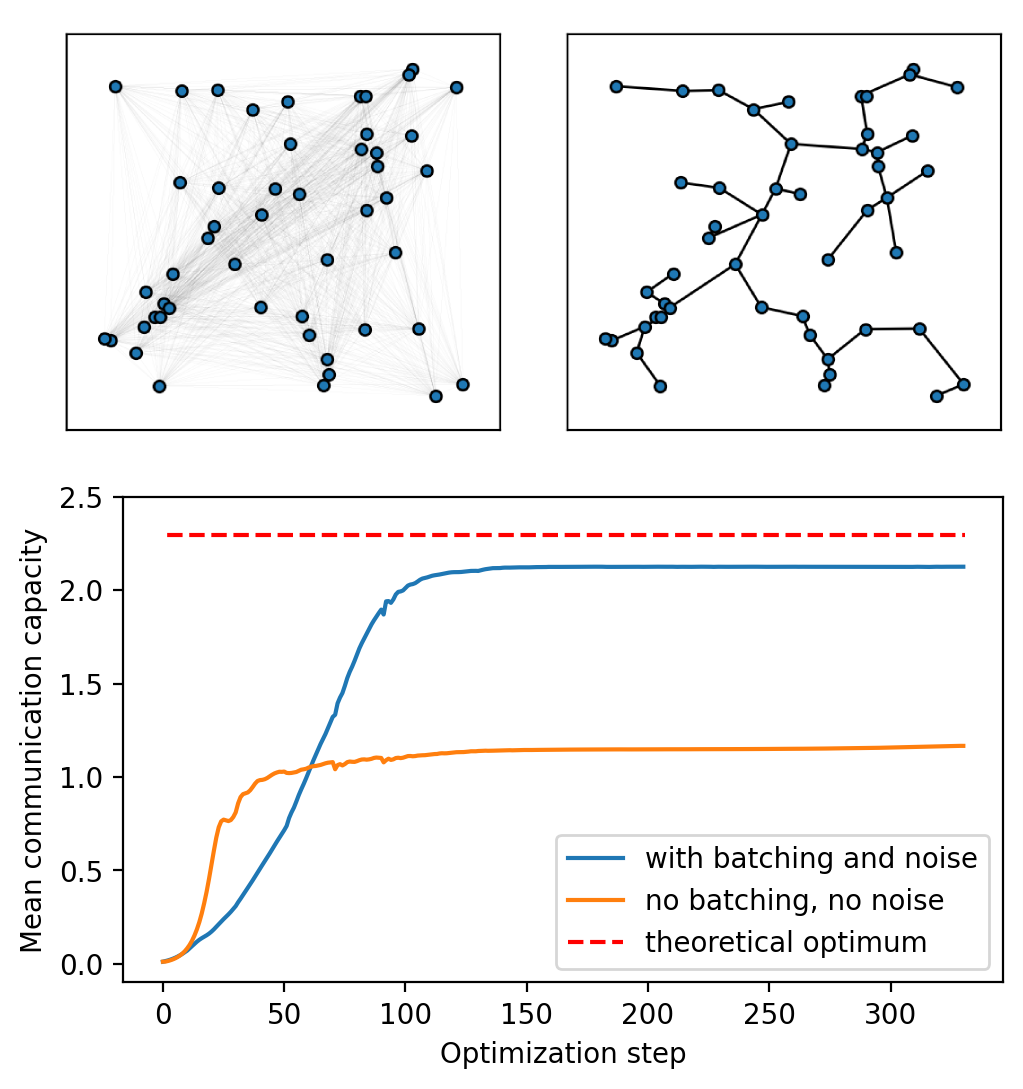}
		\caption{
            \textbf{Optimization of a spatial quantum network converges to the minimum spanning tree.} \textbf{(Top)} Initial dense network and the final optimized configuration obtained by maximizing the mean communication capacity $\mathcal C$ (Eq.~\eqref{quantum_network_loss}) under the geographic budget constraint (Eq.~\eqref{eq:quantum_budget}), with edge costs proportional to inter-node distance. \textbf{(Bottom)} Final objective value versus training step with and without stochastic smoothing. Plain gradient descent stalls at a clearly suboptimal configuration. Noise and batching do not drive the nested $\min$--$\max$ objective to its exact optimum, but they come close enough for the resulting structure to be recognized as the minimum spanning tree --- the observation that motivates the analytical treatment in \nameref{sec:methods}.
		}
		\label{fig:quantum_network}
	\end{figure}
	
	Next, we optimize the communication capacity of a quantum internet with lossy channels. Quantum networks aim to distribute entanglement and quantum information between spatially separated quantum processors, forming the backbone of future distributed quantum computing and secure communication infrastructures \cite{kimble2008quantum,wehner2018quantum}. The design of such networks involves nontrivial trade-offs between physical connectivity, channel losses, and resource constraints, and has therefore motivated a growing body of work on quantum network architecture and routing optimization \cite{van2016path,gyongyosi2018entanglement,pirandola2019end}. We include this example as a case study because it illustrates a distinct mechanism inducing sparsity, and edge costs proportional to edge lengths in spatial networks. It also demonstrates the need for more advanced gradient-descent methods incorporating batching and stochasticity.
    
    % \todo{i arch}
	The geographic placement of quantum computers is sampled randomly, and wiring costs are assumed to be proportional to the distance between nodes. We compute the pairwise distances $D_{i j}$, and optimize the adjacency matrix $A_{i j}$ under the constrained construction budget.
	
	\begin{equation}\label{eq:quantum_budget}
		B=\sum_{i j}D_{i j}A_{i j}.
	\end{equation}

    % \todo{ii loss function}
    
	Following the derivations in \cite{pirandola2019end}, we define the transmissivity of each edge $e$ as $\eta_e$. We model the edge transmissivities from the edge weights using the mapping $\eta_{i j} = \tanh{A_{i j}}$. The composite transmissivity along some path $P=\{e_1, e_2, ..., e_K\}$ is given by the bottleneck transmissivity along the chain $\eta_P = \min_{e\in P}\eta_e$ (see  \cite{pirandola2019end} for details). The capacity of this chain is given by
	
	\begin{equation}
		\mathcal C_P = -\log_2(1-\eta_P).
	\end{equation}
	And the communication capacity between two nodes $(i,j)$ in a network is the capacity of the best path $P_{i j}$ connecting them
	
	\begin{equation}
		\mathcal C_{i j} = \max_{P_{i j}} \mathcal C_P = -\log_2(1-\max_{P_{i j}}\min_{e\in P}\eta_e).
	\end{equation}
	The mean network capacity, and the target of our optimization, is then given by

	\begin{equation}\label{quantum_network_loss}
    \begin{split}
		\mathcal C({\mathbf A}) &= \frac{1}{N(N-1)} \sum_{i\ne j} {\mathcal C_{i j}} \\
        &= -\frac{1}{N(N-1)} \sum_{i\ne j}\log_2(1-\max_{P_{i j}}\min_{e\in P}\eta_e).
    \end{split}
	\end{equation}

    We minimize the loss that maximizes the mean network capacity $\ell(\mathbf A) = -\mathcal C({\mathbf A})$.
    
    The nested \(\max\)--\(\min\) objective is piecewise differentiable and is therefore analyzed using the active-region and tie-surface procedure; \nameref{sec:methods} carry out this analysis.

    % \todo{iv. discovery}
    Optimizing the quantum communication capacity in Eq.~\eqref{quantum_network_loss} subject to the geographically motivated budget constraint in Eq.~\eqref{eq:quantum_budget}, using the noise-and-batching procedure described above, recovers the minimum spanning tree (MST) as the optimal network architecture (Fig.~\ref{fig:quantum_network}). 
    \nameref{sec:methods} explain the analytical mechanism behind this observation. We first prove that any globally optimal positive-weight network structure must be acyclic: a maximum spanning forest constructed from the optimized edge weights preserves all pairwise bottleneck capacities, while every remaining cycle edge consumes positive budget without improving the objective. Thus, cyclic solutions are strictly dominated by deleting redundant edges and reallocating the saved budget. 
    \nameref{sec:methods} then prove a conditional result: if the optimal network structure is connected, then it must have uniform edge weights and coincide with the Euclidean MST. 
    The connectedness assumption is not guaranteed for every finite instance of the mean-capacity objective; rather, in our experiments it emerges as a statistical property of large random Euclidean networks, where the all-pairs average favors connecting nearly all nodes over concentrating capacity within a small cluster. Under this emergent connectedness, the distance-weighted budget selects the cheapest connected backbone, yielding the MST.

    % \todo{iii train}
	Because the objective combines nested $\min$ and $\max$ operations, its loss landscape is highly non-smooth and can contain numerous shallow local minima that trap standard gradient-based optimization. To mitigate this, we introduce two additional hyperparameters, \emph{noise} and \emph{batch size}, and optimize a locally averaged objective. Rather than evaluating the objective once per update at $\mathbf{P}$, we draw a batch of $b$ perturbed parameter vectors $\{\mathbf{P}_i\}_{i=1}^b$ in a neighborhood of $\mathbf{P}$ such that $ \lVert \mathbf{P} - \mathbf{P}_i \rVert_\infty \le \epsilon $, where $b$ denotes the batch size and $\epsilon$ the noise radius. We evaluate the objective at each $\mathbf{P}_i$, average these values, and then compute the gradient of this averaged objective to update $\mathbf{P}$ via gradient descent. This procedure acts as a stochastic local smoothing (mollification) of the objective. Rather than fixing $b$ and $\epsilon$, we schedule both over the course of training, beginning with a large batch and a wide noise radius and decreasing them as optimization progresses. This anneals the objective from a heavily smoothed surrogate toward the true objective, suppressing spurious local minima early while recovering sharp minima at convergence.

	\subsection*{Synchrony-Optimal Networks: Kuramoto Model}\label{sec:kuramoto}
	We now demonstrate how gradient-based network optimization can reveal structural principles governing nonlinear Kuramoto dynamics on networks \cite{kuramoto1975international}. This subsection reproduces the results reported in \cite{mikaberidze2025emergent} and serves as a canonical example of the \textsc{GradNet} framework in action. Rather than treating network topology as fixed and tuning a global coupling parameter, as is established in studies of synchronization \cite{strogatz2000kuramoto, arenas2008synchronization}, we instead optimize the network itself under an explicit resource constraint. This perspective aligns with the extensive line of research on optimizing networks for synchrony  \cite{motter2005network, donetti2005entangled, rad2008efficient, gorochowski2010evolving, skardal2014optimal, taylor2016synchronization}, while giving rise to novel network structures and fundamentally altered dynamics.

    We include this case study because it captures the full design, discovery cycle underlying \textsc{GradNet}: direct optimization of a nonlinear dynamical objective produces unexpected structural regularities, which in turn enable theoretical characterization and can be used to refine the numerical optimization.

	\begin{figure}[t]
		\centering
		\includegraphics[width=\linewidth]{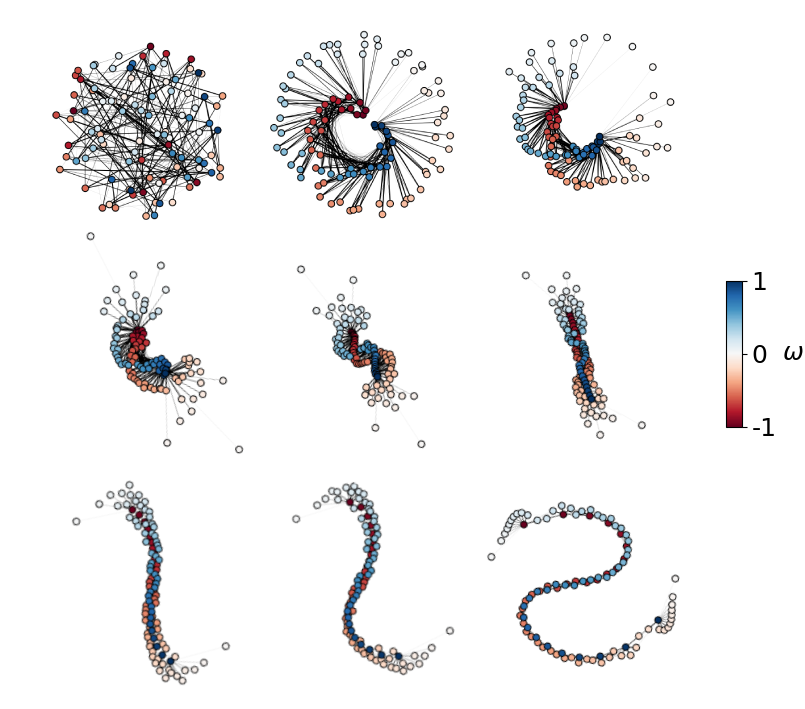}
		\caption{Sequence of snapshots of evolving network to optimize synchronization. The node colors indicate intrinsic frequencies $\omega_i$. Each subsequent snapshot supports higher value of synchronization order parameter. With this progress, the network becomes more and more sparse, bipartite (connections exist exclusively between red and blue nodes), monophilic (each node connects to a narrow range of neighboring frequencies, e.g., dark red connects to light blue nodes), and elongated (characterized by long path lengths).}
	\end{figure}

    % \todo{(i) structure}
	We focus on the Kuramoto model of heterogeneous phase oscillators,
	\begin{equation}\label{eq:kuramoto}
		\frac{d\theta_i}{dt} = \omega_i + \sum_{j=1}^N A_{i j}\sin(\theta_j-\theta_i),
	\end{equation}
	where each node has an intrinsic frequency $\omega_i$ and interacts through a weighted, undirected network encoded by the adjacency matrix $A_{i j}\ge 0$. We impose a fixed total coupling budget
	\begin{equation}\label{eq:kuramoto_budget}
		B=\sum_{i j}A_{i j},
	\end{equation}
	and search the space of undirected weighted networks with non-negative edge-weights, subject to this budget constraint. Clearly, it must obey the equimarginal principle. At the optimum, Eqs.~\eqref{eq:active_edge_condition}--\eqref{eq:inactive_edge_condition} require every positive-weight edge to provide the same marginal improvement in synchrony per unit coupling, while no absent edge can provide a larger marginal improvement.

    % \todo{(ii) loss}
    The optimization objective is to maximize collective synchrony measured as the long-time average of the Kuramoto order parameter $r(t)=\left|\tfrac{1}{N}\sum_{j=1}^N e^{i\theta_j(t)}\right|$
	\begin{equation}
		\ell(\mathbf{A})=-\frac{1}{T}\int_0^T r(t)dt
	\end{equation}
	
    % \todo{(iii) optimization} 
    We compute the gradients end-to-end through numerical integration of the dynamics using \textsc{GradNet}. Notice that each step in the gradient descent requires a new computation of the loss function and therefore a separate full integration of the dynamics.

	% \todo{(iv) discovery} 
    Despite the apparent simplicity of the objective, gradient-based optimization consistently converges to highly nontrivial network structures. The optimal networks are:
	\begin{itemize}[leftmargin=*,noitemsep]
		\item \textbf{Sparse}, with most potential edges eliminated despite no explicit sparsity penalty;
		\item \textbf{Bipartite}, connecting oscillators with frequencies of opposite sign;
		\item \textbf{Elongated}, exhibiting large diameters and long path lengths rather than small-world structure;
		\item \textbf{Extremely monophilic}, meaning that the neighbors of any node have very similar frequencies to one another while differing strongly from the node itself.
	\end{itemize}

	None of these features were imposed a priori. Instead, they emerge purely from optimization under constraints. These findings overturn several long-standing intuitions, most notably that dense, homogeneous, or short-path networks are optimal for synchronization, and demonstrate how constrained optimization can surface unexpected design principles that are difficult to anticipate analytically \cite{hong2002synchronization, nishikawa2003heterogeneity, kelly2011topology}.
	
	The emergent topology has profound dynamical implications. Optimized networks eliminate the classical synchronization threshold: partial synchrony exists for arbitrarily small per-node budgets ($b=B/N$), and global phase-locking emerges sharply once $b$ exceeds a calculable critical value. Moreover, the transition exhibits universal critical scaling distinct from that observed in fixed-topology networks.
	
	Crucially, the numerical results obtained via \textsc{GradNet} were not merely empirical. The Kuramoto model on arbitrary weighted networks is not analytically solvable in general, and even coarse characterizations of its stationary states and phase transitions typically require numerical simulation. When the network is optimized for synchrony, however, the emergent structure induces strong regularities that admit a reduced, continuum description in the thermodynamic limit \cite{mikaberidze2025emergent}.
	
	Using variational analysis, asymptotic methods, and symmetry constraints, this reduced formulation enables efficient numerical evaluation of the order parameter for arbitrary frequency distributions and yields closed-form results in important regimes, including near the phase-locking transition and in the strong-coupling limit. More broadly, this demonstrates a key principle of optimal network science: optimization can induce structural regularities that convert otherwise intractable network dynamics into mathematically structured and partially solvable problems.

    % \todo{feedback loop}
    Finally, this experiment illustrates how analytical findings can feed back into the optimization pipeline, reducing the optimization space and supplying better initializations to enable computations that would otherwise be infeasible. Specifically, gradient descent becomes unstable for per-node budget values near the critical point, making it difficult to pinpoint the global phase-locking transition numerically. Our analytical results resolve this difficulty: although the optimal edge weights depend on the budget, the optimal topology, the presence or absence of edges, does not. We can therefore take the network optimized at a larger budget and reuse it in two ways: as a mask that restricts the optimization space, and as an initial guess that stabilizes gradient descent. Together, these refinements enable accurate numerical localization of the global phase-locking transition boundary.

    \subsection*{Network Reconstruction from Functional Observables}\label{sec:net_reconstruction}

    \begin{figure}[b]
        \centering
        \includegraphics[width=0.99\linewidth]{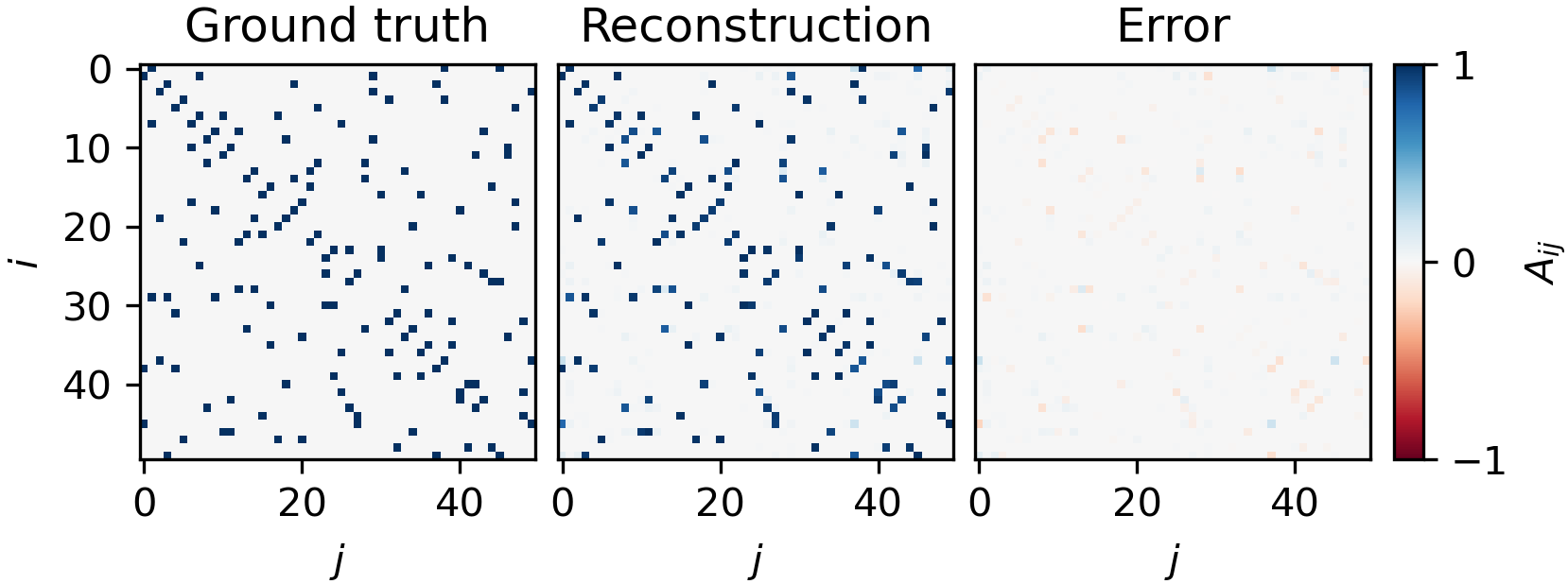}
        \caption{
            \textbf{Optimization-based network reconstruction from oscillator phase trajectories.}
            (a) Ground-truth adjacency matrix of a random 3-regular network with $N=50$ nodes used to generate Kuramoto phase trajectories $\{\bm{\theta}^{(\tau)}\}$.
            (b) Adjacency matrix reconstructed by minimizing the prediction loss defined in Eq.~\eqref{reconstruction_loss}.
            (c) Elementwise adjacency error $\mathbf A_{\mathrm{pred}}-\mathbf A_{\mathrm{true}}$; after thresholding the reconstructed edge weights at $A_{ij}=0.5$, the recovered topology exactly matches the ground truth.
        }
        \label{fig:network_reconstruction}
    \end{figure}

    The examples discussed so far use gradient-based optimization to design or modify networks for a prescribed function. We now consider the inverse problem: inferring network structure from observed dynamics. Reconstructing interactions from dynamical observations is a long-standing problem in nonlinear dynamics and network science, with applications ranging from neuroscience to social and technological systems \cite{tirabassi2015inferring, pikovsky2018reconstruction, gaskin2024inferring}. This example illustrates that the same differentiable formulation used above for network design can also be used for network inference.
    
    % \todo{(i) optimization space}
    We simulate the Kuramoto model in Eq.~\eqref{eq:kuramoto} on an unweighted random 3-regular network of $N=50$ nodes and record the phase trajectories $\{\bm\theta^{(\tau)}\}$. The intrinsic frequencies $\omega_i$ are assumed known. We reconstruct the network from scratch over the space of nonnegative, undirected weighted networks, allowing every pair of distinct nodes to interact and fixing the coupling budget in Eq.~\eqref{eq:kuramoto_budget} to that of the generating network.
    
    % \todo{(ii) loss}
    For a candidate adjacency matrix $\mathbf A$, each observed state $\bm\theta^{(\tau)}$ is used as an initial condition, and the Kuramoto dynamics are integrated over one observation interval to predict the next state $\bm{\hat{\theta}}^{(\tau+1)}$. We minimize the mean one-step phase-prediction error
    \begin{equation}\label{reconstruction_loss}
        \ell(\mathbf A)
        =
        \frac{2}{N(K-1)}
        \sum_{\tau=0}^{K-2}\sum_{i=1}^{N}
        \left[
        1-\cos\left(
        \hat{\theta}^{(\tau+1)}_i-\theta^{(\tau+1)}_i
        \right)
        \right],
    \end{equation}
    where $K$ is the number of observed states. Equivalently, this loss measures the squared distance between phases embedded on the unit circle, avoiding artificial errors due to phase wrapping.
    
    % % \todo{(iii) optimization}
    At every gradient-descent step, the one-step predictions are recomputed for the current network and gradients are propagated through the numerical integration. Minimizing Eq.~\eqref{reconstruction_loss} therefore reallocates the fixed coupling budget toward interactions that best reproduce the observed dynamics.
    
    % \todo{(iv) analysis}
    As shown in Fig.~\ref{fig:network_reconstruction}, \textsc{GradNet} recovers both the connectivity pattern and uniform edge weights of the generating network. Although no explicit sparsity penalty or prior knowledge of the 3-regular topology is imposed, the fixed coupling budget concentrates onto the true interactions. Thresholding the reconstructed edge weights at $A_{ij}=0.5$ exactly recovers the ground-truth topology. We repeated the experiment for 10 independent realizations and obtained exact topology reconstruction in every case. Thus, dynamical observations can define a loss landscape whose optimization reveals the sparse network structure underlying the dynamics.
	
    The budget plays a different role here than above: not a resource to be spent, but an assumption about the generating system that restricts the optimization space. The equimarginality condition also manifests differently: Since the true network attains $\ell=0$, the marginal gains across all edges become zero. Thus equimarginality is degenerate and becomes simply $\nabla_{ij}\ell=0$.

    \section*{Discussion}
    
    \textsc{GradNet} represents a differentiable framework for designing and studying functionally optimal networks: a scalar objective, a constrained parameterization of admissible networks, gradient-based search, and structural/dynamical analysis of the result (Fig.~\ref{fig:gradient_descent}). This complements the dominant paradigm in network science, which studies dynamics on networks that are either measured or drawn from generative models imposing structural properties a priori \cite{erdos1959random, watts1998collective, barabasi1999emergence}. In either case, the network properties are assumed, not explained. Rather than asking which dynamics a given structure supports, \textsc{GradNet} inverts the question: what structures emerge to best enable a specified function under given constraints? This positions \textsc{GradNet} in a dual role, as a scalable engineering tool for numerical network design and as a scientific probe for uncovering structure--function relationships that would be difficult to anticipate analytically.
    
    In our case studies, several canonical network features emerge from the interplay of optimization objectives and constraints: sparsity, bipartition, monophily, spanning tree backbone, community structure, and lattice symmetry. Such in silico experiments enable the rare leverage to study reasons why these canonical features might abound in the real systems. For example, sparsity can be induced by bounded message-passing depth (logical-gate networks) and competition of marginal gains (Kuramoto dynamics and quantum networks), among other mechanisms. Minimum spanning trees may emerge when pairwise path bottlenecks are maximized and redundant paths do not matter (quantum internet). Community structures can emerge when homophilic interactions are preferred and edge weights are restricted to remain non-negative (Zachary's Karate Club). Bipartite structures may arise from strong preference for disassortative links with limited node degrees (see Kuramoto model). Read together, these results suggest that structural patterns long treated as empirical descriptors of real-world networks may be understood as signatures of constrained optimality rather than as primitives to be assumed.
    
    A second theme is that optimal networks can be analytically more tractable than generic ones, precisely because optimality is a strong mathematical condition. Marginal-return equalization and the associated Lagrangian stationarity conditions turn numerically optimized structures into candidates for closed-form analysis, often by first exposing a symmetry that reduces the effective number of degrees of freedom. The lattice case yields an exact expression for the optimal edge weights, Eq.~\eqref{eq:l2_optimal_edgeweights_result}, despite closed-form Laplacian spectra being unavailable for square lattices with arbitrary weights; the quantum-internet case admits a proof that connected optima have uniform weights and coincide with the Euclidean MST; and the synchrony-optimal Kuramoto networks induce regularities that support a reduced continuum description in the thermodynamic limit, yielding closed-form results near the phase-locking transition and in the strong-coupling limit \cite{mikaberidze2025emergent}. More broadly, constrained optimization does not merely locate a good network, it can convert otherwise intractable dynamics on general weighted graphs into structured, solvable problems. Furthermore, optimizing the network structure can change the qualitative collective dynamics, such as eliminating and introducing phase transitions and producing critical scaling distinct from that of fixed-topology networks. Thus optimized structures may improve analytical tractability and support genuinely novel dynamical regimes.
    
    Methodologically, \textsc{GradNet} repurposes the automatic-differentiation and optimization infrastructure built for training artificial neural networks \cite{baydin2018automatic}, but its object of interest is different: rather than tuning the weights of a fixed architecture to minimize a task loss, \textsc{GradNet} treats topology itself as the trained object, and the resulting network structure and its dynamics --- not predictive performance --- are the outcome of interest. Within this scope, the framework unifies three problem types that are typically treated separately: construction from scratch, modification of an existing network, and inference of structure from observed dynamics (Table~\ref{tab:case_studies_summary}), under static or dynamic objectives, discrete- or continuous-time objectives, and across a range of structural constraints on directedness, edge signs, feasibility masks, and construction budgets. Relative to prior network-optimization work, this is broader than spectral-surrogate methods \cite{pecora1998master, wang2003epidemic, donetti2006optimal}. This is also more scalable than combinatorial rewiring/annealing schemes \cite{schneider2011mitigation, donetti2005entangled, rad2008efficient} that are typically suitable for $10^3$--$10^4$ nodes, and reinforcement-learning approaches \cite{zhu2021network, li2023network} that are typically limited to $10$--$100$ nodes. Differentiating directly through the objective and through simulated dynamics, together with modern optimizers, GPU execution, and sparse-tensor encoding, allows \textsc{GradNet} to scale past $10^5$ nodes (Fig.~\ref{fig:l2_variants}), with the achievable scale depending on the sparsity of the admissible set and the cost of evaluating the objective.
    
    These advantages come with corresponding limitations. The framework requires a differentiable --- or at worst piecewise-differentiable --- objective; non-smooth objectives involving shortest paths, bottlenecks, or $\min$/$\max$ operations, as in the quantum-internet case, require stochastic mollification via noise and batching during optimization, introducing additional hyperparameters that must be tuned. Gradient descent offers no general guarantee of reaching a global optimum in these nonconvex landscapes. \textsc{GradNet}'s native output is a weighted network, so recovering an unweighted topology requires an added thresholding step (Fig.~\ref{fig:network_reconstruction}). Sign constraints on modifications and on final edge weights can be jointly infeasible; \textsc{GradNet} detects and reports but does not resolve such conflicts. Finally, the current framework optimizes edge weights and topology only --- node-level properties, such as the number of nodes or fixed parameters like oscillator frequencies, are treated as given rather than as objects of optimization.
    
    Several extensions follow naturally: Multiobjective formulations for networks balancing competing objectives, temporal and multilayer networks, higher-order interactions, robustness to uncertainty in the constraints or the objective, and discrete combinatorial topology beyond the continuous relaxation used here. A further, complementary direction concerns inference rather than design: if optimality is a generative mechanism for network structure, can we infer the objectives and constraints that produced observed real-world networks? More broadly, we suggest that treating network topology as a continuously differentiable, variationally optimized object extends the list of organizing principles in Table~\ref{tab:variational_frameworks} to network science, offering a common language for design, inference, and mechanistic explanation across physical, biological, and technological systems.

    \section*{Materials and Methods}\label{sec:methods}
    
    \begin{figure*}[!t]
        \centering
        \includegraphics[width=\textwidth]{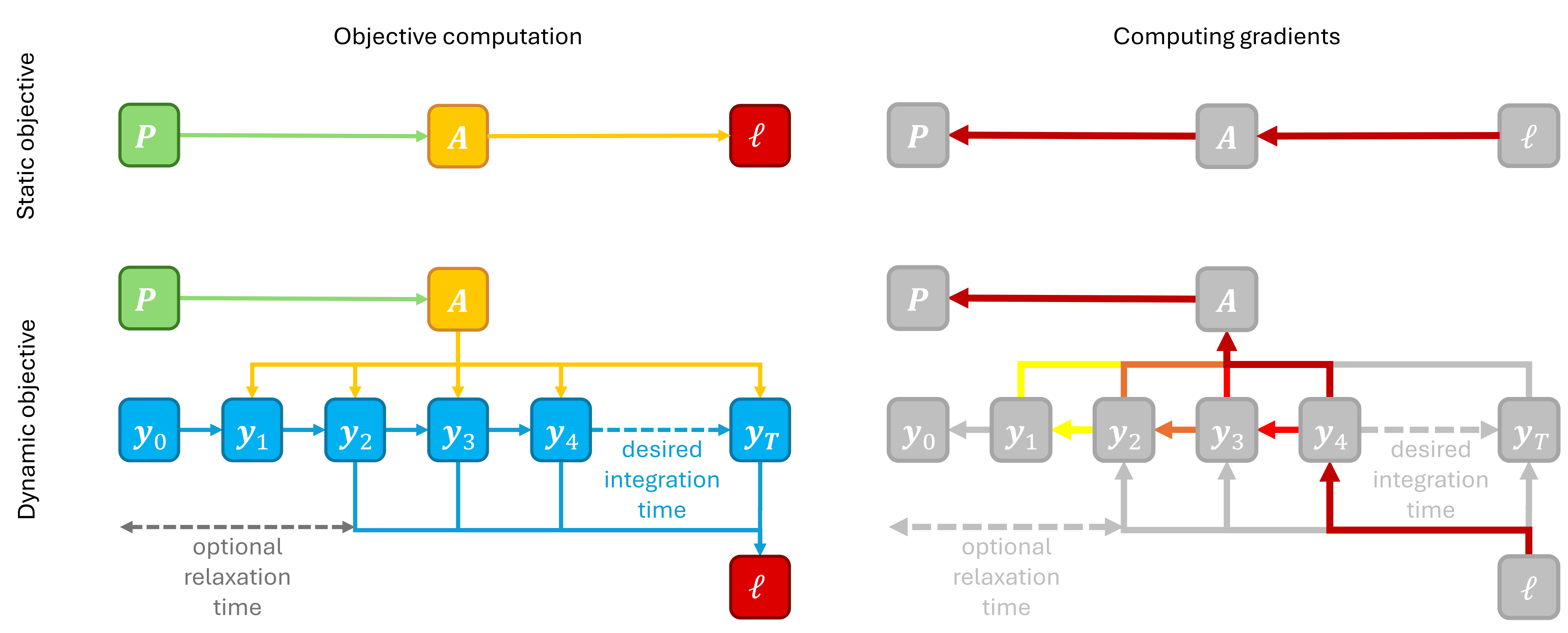}
        \caption{
        \textbf{Objective evaluation and gradient computation in \textsc{GradNet}.}
            \textbf{Top:} For a static objective, the loss is computed directly from the adjacency matrix $\mathbf A$.
            \textbf{Bottom:} For a dynamic objective, $\mathbf A$ participates in the time evolution of the dynamical state $\bm y_t$, and the loss may depend on the final state, the full trajectory, or selected states after an optional relaxation period. The right panels illustrate differentiation through these computations. The highlighted paths show the contributions to $\partial\ell/\partial P_{ij}$ arising from the chain-rule term $\partial\ell/\partial \bm y_4 \cdot d\bm y_4/dP_{ij}=\sum_{abc}\partial\ell/\partial y_{4,a} \cdot d y_{4,a}/dA_{bc} \cdot dA_{bc}/dP_{ij}$, corresponding to the backpropagation paths through $\bm y_4$. The total derivative is obtained by summing the contributions from all such paths via the chain rule. For continuous-time dynamics, the displayed states are estimated values produced by numerical integration (e.g., forward Euler).
        }
        \label{fig:architecture}
    \end{figure*}

    \subsection*{Parameterization of the Optimization Space}\label{sec:param}

    We now describe the encoding pipeline that maps $\mathbf{P} \in \mathbb{R}^{N \times N}$ to $\mathbf A\in\mathcal{A}$. We first present this pipeline in its dense form; afterward, we discuss how it can be reduced from $O(N^2)$ to $O(E)$ parameters where $E$ is the number of admissible edges when the network is \emph{a priori} known to be sparse.
	
	\textbf{Step 1: Directed/undirected networks.} 
	In many applications, the network is fundamentally undirected, and therefore its adjacency matrix must be symmetric. We enforce this structural constraint by symmetrizing the parameter matrix:
	\begin{equation}
	    \bm S
    	=
    	\begin{cases}
    		\mathbf P, 
    		& \text{if the network is directed},\\[6pt]
    		\frac{1}{2}\bigl(\mathbf P + \mathbf P^\top\bigr), 
    		& \text{if the network is undirected}.
    	\end{cases}
	\end{equation}
	
	\textbf{Step 2: Modification sign constraints.} 
	Depending on whether edge weights are allowed to increase (e.g., build or reinforce), decrease (e.g., demolish or weaken), or change in either direction, we impose sign constraints on the modifications:
    \begin{equation}
        T_{i j}
    	=
    	\begin{cases}
    		S_{i j}, 
    		& \text{if arbitrary changes are allowed},\\[6pt]
    		S_{i j}^2, 
    		& \text{if only nonnegative changes are allowed},\\[6pt]
    		-S_{i j}^2, 
    		& \text{if only nonpositive changes are allowed}.
    	\end{cases}
    \end{equation}
	
	\textbf{Step 3: Edge masking.} 
	Certain edges may be infeasible to construct or modify due to physical, geometric, or policy constraints. These restrictions are encoded in a mask matrix $\bm M$, where $M_{i j} = 1$ if the edge $(i,j)$ is allowed and $M_{i j} = 0$ otherwise. If no mask is provided, we assume that all edges except self-loops are admissible and initialize
	\begin{equation}
	    M_{i j} = 1 - \delta_{i j},
	\end{equation}
	where $\delta_{i j}$ denotes the Kronecker delta.
	
	We then eliminate forbidden modifications via element-wise multiplication:
	\begin{equation}
	   \bm U = \bm T \odot \bm M.
	\end{equation}
	This operation sets all disallowed edge modifications to zero while retaining the admissible ones.

	\textbf{Step 4: Budget constraint.} 
	We now constrain the magnitude of the modification to satisfy the allocated budget $B$. The budget is enforced using a cost matrix $\bm C$ together with an aggregation norm $p$. The parameter $p$ determines the elementwise $\ell_p$ norm used to aggregate costs across edges (for example, $p=2$ corresponds to the Frobenius norm). The entry $C_{i j}$ represents the cost of changing the edge weight between nodes $(i,j)$ by one unit. These costs may vary across node pairs, for example, they may be proportional to geometric distance. 
	
	If no cost matrix is provided, we assume uniform costs and set $C_{i j}=1$ for all admissible edges. The default aggregation norm is $p=1$, corresponding to a simple sum of costs per edge.	Next, we compute the scaling factor
    \begin{equation}
    	\gamma
    	=
    	\frac{B}{\|\bm C \odot \bm U\|_p}.
    \end{equation}
	This scalar rescales the proposed modification so that the total cost matches the prescribed budget. Notice that for undirected networks, each edge contributes to the total budget twice. We adopt this convention so that the same budget equation applies to both directed and undirected networks. Depending on the problem formulation, the budget may be enforced exactly, imposed as an upper bound, or omitted entirely:
    \begin{equation}
    	\bm \Delta
    	=
    	\begin{cases}
    		\gamma \, \bm U,
    		& \text{if $B$ is enforced exactly},\\[6pt]
    		\min(\gamma,1)\, \bm U,
    		& \text{if $B$ is an upper bound},\\[6pt]
    		\bm U,
    		& \text{if there is no budget constraint}.
    	\end{cases}
    \end{equation}

	\textbf{Step 5: Building or modifying the adjacency matrix.} 
	With $\bm \Delta$ determined, we construct the updated adjacency matrix as
    \begin{equation}
    	\tilde{\mathbf A} = \mathbf A^{(0)} + \bm \Delta.
    \end{equation}
	
	Here $\mathbf A^{(0)}$ denotes the initial adjacency matrix prior to optimization. If an initial network is provided, $\bm \Delta$ represents modifications to that baseline structure. If no initial network is specified, we set
    \begin{equation}
        \mathbf A^{(0)} = 0,
    \end{equation}
	which corresponds to constructing the network entirely from scratch.

	\textbf{Step 6: Edge weight sign constraints.} 
	Finally, we enforce the prescribed sign constraints on the edge weights to obtain the final adjacency matrix:
    \begin{equation}
	\mathbf A
	=
	\begin{cases}
		\tilde{\mathbf A},
		& \text{if edge weights are unconstrained},\\[6pt]
		|\tilde{\mathbf A}|_\epsilon,
		& \text{if edge weights must be nonnegative},\\[6pt]
		-|\tilde{\mathbf A}|_\epsilon,
		& \text{if edge weights must be nonpositive}.
	\end{cases}
    \end{equation}
	
	Here $|\cdot|_\epsilon$ denotes a smooth approximation of the absolute value, introduced to avoid non-differentiable kinks in the loss landscape. For a scalar $x$, it is defined as
    \begin{equation}
        |x|_\epsilon = x \tanh\!\left(\frac{x}{\epsilon}\right),
    \end{equation}
	where $\epsilon \ll 1$ is a small smoothing parameter. Thus, the overall differentiable encoding $g:\mathbb{R}^{N\times N}\to\mathcal{A}$, satisfying $\mathbf A=g(\mathbf P)$, is the composition of the transformations described above.

	It is important to note that when the sign constraints imposed on the modifications $\bm \Delta$ conflict with the sign constraints imposed on the final edge weights $\mathbf A$, the pipeline may not simultaneously satisfy the modification-sign constraint and exact budget exhaustion. In such cases, strict adherence to both conditions cannot be guaranteed. To avoid ambiguity, the \textsc{GradNet} Python package automatically detects and reports any such violations to the user. Moreover, if the network is undirected, then $\mathbf A^{(0)}$, $\bm M$, and $\bm C$ must be symmetric. This condition is verified and, when necessary, enforced by \textsc{GradNet} to ensure consistency of the optimization space.
	
    \textbf{Possible sparse encoding.} 
	Algorithmically, the \textsc{GradNet} optimization framework can be implemented with matrices that are either dense or sparse encoded. When the network is known \emph{a priori} to be sparse, meaning that most edges are infeasible and the mask matrix $\bm M$ is largely zero, it is inefficient to store and manipulate matrices in dense form. For such cases, the \textsc{GradNet} Python package supports encoding all matrices as sparse \textsc{PyTorch} tensors, which is triggered automatically whenever the user provides a sparsely encoded mask $\bm M$. The entire encoding pipeline is then executed in sparse form, requiring $O(E)$ memory, where $E$ denotes the number of admissible edges, instead of $O(N^2)$. This also avoids unnecessary subsequent dense matrix operations in the loss computation, resulting in substantial savings in both memory usage and computational cost.

    \subsection*{Loss Function Definition}\label{sec:loss}

    The user supplies the second map, $\ell:\mathcal A\to\mathbb R$, as a differentiable Python function. \textsc{GradNet} supplies the encoding $\mathbf A=g(\mathbf P)$, composes it with the loss function, and uses automatic differentiation to minimize the resulting objective. Objectives to be maximized are implemented by negating the corresponding performance measure.
    
    We call an objective \emph{static} when the loss is computed directly from the network structure, without simulating an evolving state. Examples include functions of the adjacency or Laplacian spectrum, such as the algebraic connectivity. The calculation may still involve iterative numerical algorithms.
    
    A dynamic objective depends on a process unfolding on the network. In discrete time, the state may evolve as $\bm y_{t+1}=F_t(\bm y_t,\mathbf A)$. After evolution to time $T$, the loss may depend on $\bm y_T$, the full trajectory $\{\bm y_t\}$, or a selected time interval, such as the states following
    an initial relaxation period. It may also be averaged over multiple initial conditions or inputs. 
    
    For continuous-time dynamics, the states in Fig.~\ref{fig:architecture} are generated by a numerical integration scheme. Higher-order methods such as Runge--Kutta introduce additional intermediate calculations, but these remain part of the same differentiable computation from $\mathbf A$ to $\ell$.
    
    \textsc{GradNet} utilizes PyTorch and torchdiffeq \cite{paszke2019pytorch, torchdiffeq} to automatically apply the chain rule through the encoding, dynamics, and loss calculation and compute $\partial\ell/\partial P_{ij}$. The user therefore does not need to derive these gradients manually. The loss must, however, be implemented using PyTorch tensor operations with defined derivatives. Dynamic objectives additionally require a new simulation each time the loss is evaluated.

    \subsection*{Gradient Descent Optimization}\label{sec:grad}
    
	Using standard gradient descent, the parameters are updated according to	
	\begin{equation}\label{gradient_descent}
		P^{(n+1)}_{i j}
		=
		P^{(n)}_{i j}
		-
		\eta \frac{\partial \ell(\mathbf A)}{\partial P_{i j}}.
	\end{equation}
	Here $\eta$ denotes the learning rate, which determines the step size taken in the parameter space at each iteration. \textsc{GradNet} Python package allows usage of arbitrary PyTorch optimizers and learning-rate schedules.

	\textbf{Initialization.} 
	Depending on the problem type, the parameters $P_{i j}$ may be initialized randomly, uniformly, or using a mixture of both. The choice of initialization can significantly affect optimization stability and convergence behavior, and different problem classes may favor different strategies.

    \textbf{Normalization.} 
	We further stabilize the optimization process by observing that, when a budget constraint is enforced, the free parameters $\mathbf P$ may be rescaled without altering the encoded adjacency matrix $\mathbf A$. Although this scaling invariance leaves the forward map $\mathbf A = g(\mathbf P)$ unchanged, it does influence gradient magnitudes and therefore affects optimization dynamics. Empirically, we find that stability improves when $\mathbf P$ is renormalized after each gradient step to satisfy
	\begin{equation}
	    \|\mathbf P\|_2 = \sqrt{E},
	\end{equation}
	where $E$ denotes the number of unmasked (i.e., admissible) edges, corresponding to the effective number of degrees of freedom. This renormalization is enabled by default in the Python implementation of \textsc{GradNet}.
    %, but may be disabled by the user if desired.

    \textbf{Stochastic Gradient Descent.} 
	In problems where the loss landscape contains many local minima, it can be beneficial to introduce stochastic smoothing to reduce the likelihood of becoming trapped. In \textsc{GradNet}, this is implemented by injecting noise into the parameters $\mathbf P$, thereby producing perturbed adjacency matrices $\mathbf A$ in a neighborhood of the current iterate. 
    Noise can be additive $P_{i j}\mapsto P_{i j} +\epsilon_{i j}$ or multiplicative $P_{i j}\mapsto (1+\epsilon_{i j})P_{i j}$. A batch of such perturbations is sampled, the loss is evaluated at each sample, and the gradient is computed with respect to the averaged objective. This procedure acts as a local smoothing of the loss landscape and can substantially improve robustness. As explored in ``\nameref{sec:quantim_internet}'' case study.
	
    \textbf{Alignment with ML infrastructure.} 
	A key advantage of this framework is its seamless compatibility with modern machine learning infrastructure. Instead of the basic gradient descent update in Eq.~\eqref{gradient_descent}, one may employ advanced optimizers such as \textit{Adam}, incorporate learning-rate schedules, execute computations on GPUs, and monitor training dynamics in real time using \textit{TensorBoard}. This integration allows network optimization problems to directly leverage the efficiency and robustness of contemporary differentiable programming tools.

    \subsection*{Optimal lattice for $\lambda_2$} \label{laplacian_appendix}
    \subsubsection*{Algorithm for computing $\lambda_2$ in sparse encoding}
    Many standard algorithms for matrices rely on their dense encoding. Here we present a method of estimating the second eigenvalue of a sparsely encoded Laplacian matrix. Since the standard power method returns the eigenvalue of largest magnitude, whereas $\lambda_2$ is the second-smallest Laplacian eigenvalue, we instead consider the shifted matrix
	\begin{equation}
		\tilde L_{i j} = c \delta_{i j} - L_{i j}.
	\end{equation}
	Its eigenvalues satisfy $\tilde\lambda_i = c - \lambda_{N+1-i}$, where $N=n^2$ is the network size and hence the dimension of the Laplacian matrix. By Gershgorin’s circle theorem \cite{gershgorin1931uber}, the Laplacian eigenvalues are bounded above by $2\max_i L_{ii}$; choosing $c = 2\max_i L_{ii}$ ensures $\tilde\lambda_i \ge 0$, so they are ordered by magnitude.
	
	The dominant eigenvector of $\tilde{\mathbf{L}}$ corresponds to the constant eigenvector $\bm v_1 = \bm 1 / \sqrt{N}$ of $\mathbf{L}$. We therefore initialize with a random vector $\bm r$ and project out this component,
	\begin{equation}
		\bm x^{(0)} = \bm r - (\bm r^\top \bm v_1)\bm v_1,
	\end{equation}
	and iterate
	\begin{equation}
		\bm x^{(k+1)} = \tilde{\mathbf{L}}\bm x^{(k)}.
	\end{equation}
	This iteration converges to $\lim_{k\to\infty}\bm x^k = \bm{\tilde v}_{N-1} = \bm v_2$, 
    % \todo{how define converge... w/ threshold?} 
    from which we compute $\lambda_2$ via the Rayleigh quotient,
	\begin{equation}\label{rayleigh_quotient}
		\lambda_2 = \frac{\bm v_2^\top \mathbf{L} \bm v_2}{\bm v_2^\top \bm v_2}.
	\end{equation}
	
	This procedure relies only on matrix--vector multiplications, which are efficient for sparse matrices. Differentiating only through the Rayleigh quotient in Eq.~\eqref{rayleigh_quotient} enables efficient optimization of edge weights.

    \subsubsection*{Derivation of optimal edge weights}
    Let us first express the Laplacian matrix for the $n \times n$ square lattice. The nodes are indexed by $i$ and $j$ to indicate the vertical and horizontal positions respectively. We denote the horizontal edge weight connecting to the node $(i,j)$ from right as $a_{i j}$. Similarly, the vertical edge weight connecting from below is denoted $b_{i j}$. Since we use two indices per node, the state vector $x_{i j}$ also has two indices. In these terms, the action of the Laplacian operator $L_{ij;\alpha\beta}$ on the state $x_{i j}$ can be expressed as
	
	\begin{equation}\label{laplacian_action}
	\begin{split}
		(\mathbf{L} \bm x)_{i j} = & \sum_{\alpha\beta} L_{ij;\alpha\beta} x_{\alpha\beta}
		\\
		= & a_{i,j-1}(x_{i j} - x_{i,j-1}) + a_{i j}(x_{i j} - x_{i,j+1})
		\\
		+ & b_{i-1,j}(x_{i j} - x_{i-1,j}) + b_{i j}(x_{i j} - x_{i+1,j})
	\end{split}
	\end{equation}
	
	The first eigenvector of the Laplacian for a connected network is uniform $\bm x^{(1)}=\mathds{1}$. With this, we can write down the variational characterization of the second eigenvalue
	
	\begin{equation} \label{l2_variational}
	\begin{split}
		\lambda_2&=\min_{\substack{\|\bm x\|=1 \\ \bm x \perp \mathds{1}}} \bm x^\top \mathbf{L} \bm x
		\\
		&=\min_{\substack{\|\bm x\|=1 \\ \bm x \perp \mathds{1}}} 
		\sum_{i j} \big(a_{i j}(x_{i j}-x_{i,j+1})^2 + b_{i j}(x_{i j}-x_{i+1,j})^2\big).
	\end{split}
	\end{equation}
    Here, the result on the second line is obtained by using Eq.~\eqref{laplacian_action} and simple algebraic transformations. Assuming that $\lambda_2$ is simple, the equimarginal principle together with the budget constraint Eq.~\eqref{laplacian_budget} requires the marginal gains of the performance target with respect to each of the positive edge weights to be equal:
    
	\begin{equation} \label{laplacian_eigenvectors}
	\begin{split}
		\frac{\partial{\lambda_2}}{\partial a_{i j}} = \frac{\partial{\lambda_2}}{\partial b_{i j}}  &= \text{const. (independent of i and j)}
		\\ &
		=\left(x^{(2)}_{i+1,j} - x^{(2)}_{i j}\right)^2
		=\left(x^{(2)}_{i,j+1} - x^{(2)}_{i j}\right)^2.
	\end{split}
	\end{equation}
	Here $\bm x^{(2)}$ is the minimizer of Eq.~\eqref{l2_variational}, i.e., the second eigenvector. Using these conditions alongside with the condition $\bm x \perp \mathds{1}$ we find two solutions for the second eigenvector, indicating that the second eigenvalue is degenerate
	\begin{equation}\label{eq:fiedler_evecs}
		x^{(2)}_{i j} \propto i-j    \quad \text{and} \quad    x^{(2)}_{i j} \propto i+j-n-1.
	\end{equation}
	 
    Next, we plug the eigenvectors into $\mathbf{L}\bm x = \lambda_2 \bm x$. Using this with Eqs.~\eqref{laplacian_action} and \eqref{laplacian_eigenvectors} we get the two following conditions
	\begin{equation}
		\begin{split}
			\lambda_2(i-j) &= b_{i-1,j}-b_{i j} - a_{i,j-1} + a_{i j}
			\\
			\lambda_2(i+j-n-1) &= b_{i-1,j}-b_{i j} + a_{i,j-1} - a_{i j}.
		\end{split}
	\end{equation}
	Solving this system of equations gives us $a_{ik} = b_{kj} = \frac{\lambda_2}{2} k(n-k)$ which confirms that the vectors in Eq.~\eqref{eq:fiedler_evecs} are indeed eigenvectors. Last missing piece, $\lambda_2$, can be obtained by applying the budget condition Eq.~\eqref{laplacian_budget}, resulting in $\lambda_2=\tfrac{3B}{n^2(n^2-1)}$. This gives us the closed-form expression for the optimal edge weights
	\begin{equation} \label{l2_optimal_edgeweights}
		a_{ik}=b_{kj} = \frac{3}{2}\frac{B}{n^2(n^2-1)} k(n-k).
	\end{equation}

    This captures the numerically observed property that vertical/horizontal edge weights only depend on the second/first index, and it exhibits close agreement with numerical optimization results, as demonstrated in Fig.~\ref{fig:l2_variants}.

    The derivation above has ignored that at the optimum the eigenvalue is degenerate and hence we can not take the derivatives. Still, we will treat the obtained result as an ansatz and will verify its validity. Let $c = \frac{3B}{2n^2(n^2-1)}$, so that Eq.~\eqref{l2_optimal_edgeweights} reads $a_{ik} = b_{kj} = c\,k(n-k)$. We will show that $2c$ bounds $\lambda_2$ from above for every feasible lattice and that Eq.~\eqref{l2_optimal_edgeweights} attains it.

Let $x_{ij} = i-j$. Then $\bm x \perp \mathbf{1}$, every edge difference equals $\pm 1$, and $\|\bm x\|^2 = n^2(n^2-1)/6$. For any nonnegative weights satisfying Eq.~\eqref{laplacian_budget}, i.e.\ $\sum_{ij}(a_{ij}+b_{ij}) = B/2$, Eq.~\eqref{l2_variational} gives the upper bound
\begin{equation}
  \lambda_2 \le \frac{\bm x^\top \mathbf L \bm x}{\|\bm x\|^2}
  = \frac{\sum_{ij}(a_{ij}+b_{ij})}{\|\bm x\|^2}
  = \frac{3B}{n^2(n^2-1)} = 2c.
\end{equation}

Finally, we show this bound can be attained. Under Eq.~\eqref{l2_optimal_edgeweights}, horizontal weights depend only on the column and vertical weights only on the row, so $\mathbf L = \mathbf L_P \otimes \mathbf I + \mathbf I \otimes \mathbf L_P$, where $\mathbf L_P \in \mathbb{R}^{n\times n}$ is the Laplacian of an $n$-node path with weights $w_k = c\,k(n-k)$:
\begin{equation}
  (\mathbf L_P \bm v)_k = w_{k-1}(v_k - v_{k-1}) + w_k(v_k - v_{k+1})
\end{equation}
with $k = 1,\dots,n$. Since $w_0 = w_n = 0$, this formula also holds at the endpoints. For $m = 0,\dots,n-1$, let $\bm p^{(m)} \in \mathbb{R}^n$ be the vector with entries $p^{(m)}_k = k^m$; these form a basis of $\mathbb{R}^n$. For $\bm v = \bm p^{(m)}$, it's easy to verify that the right-hand side above is a polynomial in $k$ whose $k^{m+1}$ terms cancel and whose $k^m$ coefficient is $c\,m(m+1)$. Hence
\begin{equation}
  \mathbf L_P\, \bm p^{(m)} = c\,m(m+1)\, \bm p^{(m)} + \sum_{l<m} \alpha_{ml}\, \bm p^{(l)} ,
\end{equation}
so $\mathbf L_P$ is triangular in this basis, with eigenvalues $c\,m(m+1)$, $m = 0,\dots,n-1$. The eigenvalues of $\mathbf L$ are the sums $c\,[m(m+1) + m'(m'+1)]$, so its smallest nonzero eigenvalue is $\lambda_2 = 2c$, attained with multiplicity two ($m,m' = 1,0$ and $0,1$), consistent with Eq.~\eqref{eq:fiedler_evecs}. This shows that the upper bound is saturated by Eq.~\eqref{l2_optimal_edgeweights} and therefore it is a global maximizer of $\lambda_2$, with $\lambda_2 = 3B/(n^2(n^2-1))$.

    \subsection*{Tree structure of the optimal quantum network}
    \label{quantum_appendix}
    
    We prove two structural facts about the optimization problem in ``\nameref{sec:quantim_internet}'' case study. First, we write edges as unordered pairs $e=\{i,j\}$, with length $d_e=D_{ij}$ and weight $a_e=A_{ij}$. The budget constraint in Eq.~\eqref{eq:linear_budget} then becomes
    \begin{equation}
        \sum_e d_e a_e = B/2,
        \qquad a_e\geq 0 .
    \end{equation}
    We assume $d_e>0$, and define the network structure to be the set of edges with $a_e>0$. Let
    \begin{equation}
        \phi(x)=-\log_2(1-\tanh x).
    \end{equation}
    Since $\tanh x$ is strictly increasing, maximizing the capacity in Eq.~\eqref{quantum_network_loss} is equivalent to maximizing
    \begin{equation}
        \sum_{i<j}\phi(s_{ij}),
        \qquad
        s_{ij}=\max_{P_{ij}}\min_{e\in P_{ij}} a_e ,
    \end{equation}
    where $s_{ij}$ is the widest-path bottleneck between nodes $i$ and $j$. If no path connects $i$ and $j$, we set $s_{ij}=0$.
    
    First, we show that the optimal network structure is a forest, i.e., it contains no cycles. Consider any feasible network and let $T$ be its maximum spanning forest. By the widest-path property of a maximum spanning forest \cite{hu1961maximum}, $T$ contains all pair-wise maximum-bottleneck paths. Thus, for every pair of nodes in the same connected component,
    \begin{equation}
        s_{ij}
        =
        \min_{e\in P^T_{ij}}a_e ,
    \end{equation}
    where $P^T_{ij}$ is the unique path between $i$ and $j$ in $T$. Thus deleting all edges outside $T$ preserves every pairwise bottleneck $s_{ij}$, and hence preserves the objective value. If the original network structure contained a cycle, this deletion strictly reduces the budget used, because all deleted edges have positive length and positive weight. Rescaling all remaining positive weights by a common factor then restores the budget and strictly increases every nonzero bottleneck within each connected component. Since $\phi$ is strictly increasing, the objective strictly increases. Therefore no cyclic network structure can be optimal. Hence every optimal network structure is acyclic.
    
    We next show that if an optimal network structure is connected, then all of its edge weights are equal and the network structure is the Euclidean minimum spanning tree. By the result above, a connected optimal network structure must be a spanning tree; call it $T$. Fix this labeled tree and consider an ordering of its $N-1$ edge weights,
    $a_1\geq a_2\geq \cdots \geq a_{N-1}$. Reveal the edges from largest to smallest weight. When edge $r$ is revealed, it merges two components of sizes $p_r$ and $q_r$, and exactly $p_r q_r>0$ node pairs acquire bottleneck value $a_r$. Therefore, for this fixed ordering, the objective has the form
    \begin{equation}\label{eq:ordered_objective}
        \sum_{r=1}^{N-1} p_rq_r \phi(a_r).
    \end{equation}
    The function $\phi$ is strictly convex, since
    \begin{equation}
        \phi''(x)=\frac{\operatorname{sech}^2 x}{\ln 2}>0 .
    \end{equation}
    Thus, for the given ordering, the objective is strictly convex. We will see below that, after fixing this ordering, the feasible set of edge weight configurations is a simplex, and therefore the maximum of the convex objective over this simplex must occur at one of its vertices.
    
    Introduction of the ordering and rewriting the objective as Eq.~\eqref{eq:ordered_objective} is essential here. Without fixing the order of the edge weights, the bottleneck terms in the arguments of $\phi$ contain minima, and the objective is only piecewise convex, not globally convex on the original feasible set. Instead, we partition the full feasible set into ordering regions. The objective is convex on each such region, and a global maximizer must be a vertex of one of these ordered regions.
    
    Given an ordering, we now need to identify the admissible edge weight configuration simplex and its vertices. Set $a_N=0$ and define
    \begin{equation}
        y_k=a_k-a_{k+1},
        \qquad k=1,\ldots,N-1 .
    \end{equation}
    Then $y_k\geq 0$ and
    \begin{equation}
        a_r=\sum_{k=r}^{N-1} y_k .
    \end{equation}
    The budget constraint becomes
    \begin{equation}
        \sum_{r=1}^{N-1}d_r a_r
        =
        \sum_{k=1}^{N-1}
        \left(\sum_{r=1}^{k}d_r\right)y_k
        =
        \frac B 2.
    \end{equation}
    Thus, in the variables $y_k$, the ordered feasible region is an ordinary simplex. Its vertices have exactly one nonzero $y_k$. Transforming back to the $a_r$ variables, the vertices are precisely the points of the form
    \begin{equation}
        a_1=\cdots=a_k=
        \frac{B}{2\sum_{r=1}^{k}d_r},
        \qquad
        a_{k+1}=\cdots=a_{N-1}=0
    \end{equation}
    for some $k\in\{1,\ldots,N-1\}$.
    
    Since the network structure is assumed connected, every edge of the spanning tree has positive weight. Therefore the only possible vertex of this form has $k=N-1$. Hence
    \begin{equation}
        a_1=a_2=\cdots=a_{N-1}
        =
        \frac{B}{2\sum_{e\in T}d_e}.
    \end{equation}
    Thus every edge in a connected optimal network structure has the same weight.
    
    It remains to identify the tree. If all edges of $T$ have common weight
    \begin{equation}
        \alpha=\frac{B}{2\sum_{e\in T}d_e},
    \end{equation}
    then every pairwise bottleneck equals $\alpha$, and the objective is proportional to $\phi(\alpha)$. Since $\phi$ is strictly increasing, maximizing the objective is equivalent to minimizing $\sum_{e\in T}d_e$. Therefore, among connected network structures, the optimal tree is the spanning tree of minimum total geometric length, namely the Euclidean minimum spanning tree. Consequently, if the optimal network structure is connected, then it is the MST and its edge weights are uniform:
    \begin{equation}
        a_e=\frac{B}{2\sum_{f\in T_{\mathrm{MST}}}d_f},
        \qquad e\in T_{\mathrm{MST}} .
    \end{equation}

    \section*{Code Availability}
    
    \textsc{GradNet} is released as an open-source Python package. The source code is available at \url{https://github.com/mikaberidze/gradnet}. Documentation, API references, and implementations of all case studies presented in this paper are provided at \url{https://gradnet.readthedocs.io/api/gradnet.html}. The package is also available through the Python Package Index and can be installed with \texttt{pip install gradnet}.

	\section*{Acknowledgments}
    This work is supported by the U.S. National Science Foundation under Grant No. DMS-2401276. Any opinions, findings, and conclusions or recommendations expressed in this material are those of the authors and do not necessarily reflect the views of the NSF. The authors acknowledge funding from the University of Wyoming including Alumbra Inc. and the AI Matching Funds program, an AI seed grant, and support from the Center for Wildlife, Technology and Computing (WyldTech) and the Center for Quantum Information Science and Engineering. This work was also supported by the EU Horizon Europe project FORGE-AI.

    \bibliographystyle{IEEEtran}
    \bibliography{references}

\end{document}